\input psfig
\input mn.tex

\def\arcosh{\rm arcosh}
\def\arcos{\rm arcos}
\def\kpc{{\rm\,kpc}}

\def\rt{R_{\rm t}}
\def\rr{R_{\rm r}}
\def\yr{y_{\rm r}}
\def\qm2{q^{-2}}
\def\d{{\rm d}}
\def\u{\underline}
\def\fr#1#2{\textstyle {#1\over #2}\displaystyle}

\def\pibytwo{{\textstyle {\pi \over 2} \displaystyle}}

\def\o{\overline}
\def\u{\underline}

\def\sign{\rm sign}
\def\rc{R_{\rm c}}
\def\xonet{x_{1, {\rm t}}}
\def\xtwot{x_{2, {\rm t}}}
\def\xoner{x_{1, {\rm r}}}
\def\xtwor{x_{2, {\rm r}}}
\def\yonet{y_{1, {\rm t}}}
\def\ytwot{y_{2, {\rm t}}}
\def\yoner{y_{1, {\rm r}}}
\def\ytwor{y_{2, {\rm r}}}
\def\dd{D_{\rm d}}
\def\ds{D_{\rm s}}
\def\dds{D_{\rm ds}}
\def\zd{z_{\rm d}}
\def\rone{\rm I}
\def\rtwo{\rm II}
\def\rthree{\rm III}
\def\rfour{\rm IV}
\def\ni{n_{\rm I}}
\def\nii{n_{\rm II}}
\def\niii{n_{\rm III}}
\def\zbar{{\o z}}
\def\Ind{\rm Ind}
\def\cw{C_\omega}
\def\cz{C_z}
\def\zs{z_{\rm s}}
\def\qn{q_{\rm n}}
\def\qu{q_{\rm u}}
\def\xic{\xi_{\rm c}}
\def\Sigmac{\Sigma_{\rm crit}}
\def\rhos{\rho_{\rm s}}
\def\rc{r_{\rm c}}
\def\zr{z_{\rm r}}
\def\zt{z_{\rm t}}
\def\lamr{\lambda_{\rm r}}
\def\mtot{{\cal M}}
%
%
\def\spose#1{\hbox to 0pt{#1\hss}}
\def\lta{\mathrel{\spose{\lower 3pt\hbox{$\sim$}}
    \raise 2.0pt\hbox{$<$}}}
\def\gta{\mathrel{\spose{\lower 3pt\hbox{$\sim$}}
    \raise 2.0pt\hbox{$>$}}}
\def\today{\ifcase\month\or
 January\or February\or March\or April\or May\or June\or
 July\or August\or September\or October\or November\or December\fi
 \space\number\day, \number\year}
\newdimen\hssize
\hssize=8.4truecm  
\newdimen\hdsize
\hdsize=16.8truecm    


\newcount\eqnumber
\eqnumber=1
\def\chaphead{}
 
\def\new{\hbox{(\rm\chaphead\the\eqnumber)}\global\advance\eqnumber by 1}
 
\def\first{\hbox{(\rm\chaphead\the\eqnumber a)}\global\advance\eqnumber by 1}
\def\last#1{\advance\eqnumber by -1 \hbox{(\rm\chaphead\the\eqnumber#1)}\advance
     \eqnumber by 1}
 
\def\ref#1{\advance\eqnumber by -#1 \chaphead\the\eqnumber
     \advance\eqnumber by #1}
 
\def\nref#1{\advance\eqnumber by -#1 \chaphead\the\eqnumber
     \advance\eqnumber by #1}

\def\eqnam#1{\xdef#1{\chaphead\the\eqnumber}}
 
 

\pageoffset{-0.85truecm}{-1.05truecm}



\pagerange{}
\pubyear{1996}
\volume{000, 000--000}


\begintopmatter

\title{Lens Models with Density Cusps}

\author{N.W.\ Evans and M.\ Wilkinson}

\vskip0.15truecm
\affiliation{Theoretical Physics, Department of Physics, 1 Keble Road,
                 Oxford, OX1 3NP}

\shortauthor{N.W.\ Evans and M.\ Wilkinson} 

\shorttitle{Lens Models with Density Cusps} 



\abstract{Lens models appropriate for representing cusped galaxies 
and clusters are developed. The analogue of the odd number theorem for
cusped density distributions is given. Density cusps are classified
into strong, isothermal or weak, according to their lensing
properties. Strong cusps cause multiple imaging for any source
position, whereas isothermal and weak cusps give rise to only one
image for distant sources. Isothermal cusps always possess a
pseudo-caustic. When the source crosses the pseudo-caustic, the number
of images changes by unity.

Two families of cusped galaxy and cluster models are examined in
detail. The double power-law family has an inner cusp, followed by a
transition region and an outer envelope. {\it One member of this
family -- the isothermal double power-law model -- possesses an
exceedingly scarce property, namely, the lens equation is exactly
solvable for any source position.} This means that the magnifications,
the time delay and the lensing cross-sections are all readily
available.  The model has a three dimensional density that is cusped
like $r^{-2}$ at small radii and falls off like $r^{-4}$
asymptotically. Thus, it provides a very useful representation of the
lensing properties of a galaxy or cluster of finite total mass with a
flat rotation curve.  The second set of models studied is the single
power-law family. These are single density cusps of infinite extent.
The properties of the critical curves and caustics and the behaviour
of the lenses in the presence of external shear are all discussed in
some detail.}


\maketitle  

\section{Introduction}

The aim of this paper is to examine the properties of lens models with
density cusps. High-resolution imaging of the the nuclei of early-type
galaxies by the Hubble Space Telescope (e.g., Lauer et al. 1995) have
provided abundant examples of cusps. In almost all early-type
galaxies, the logarithmic gradient of the surface brightness with
respect to projected radius is constant right down to the very
centre. Galaxy cores are hardly ever observed. This observational
evidence for the ubiquity of central cusps is reinforced by
theoretical suggestions of dynamical processes by which cusps can grow
in the basins of potential wells (e.g., Bahcall \& Wolf 1976; Faber et
al. 1996; Evans \& Collett 1997). Further, the influential numerical
simulations by Navarro, Frenk \& White (1996) of galaxy formation in
hierachical clustering cosmogonies found striking evidence for a
universal cusped density law for dark matter haloes and clusters.

\beginfigure*{1}

\centerline{\psfig{figure=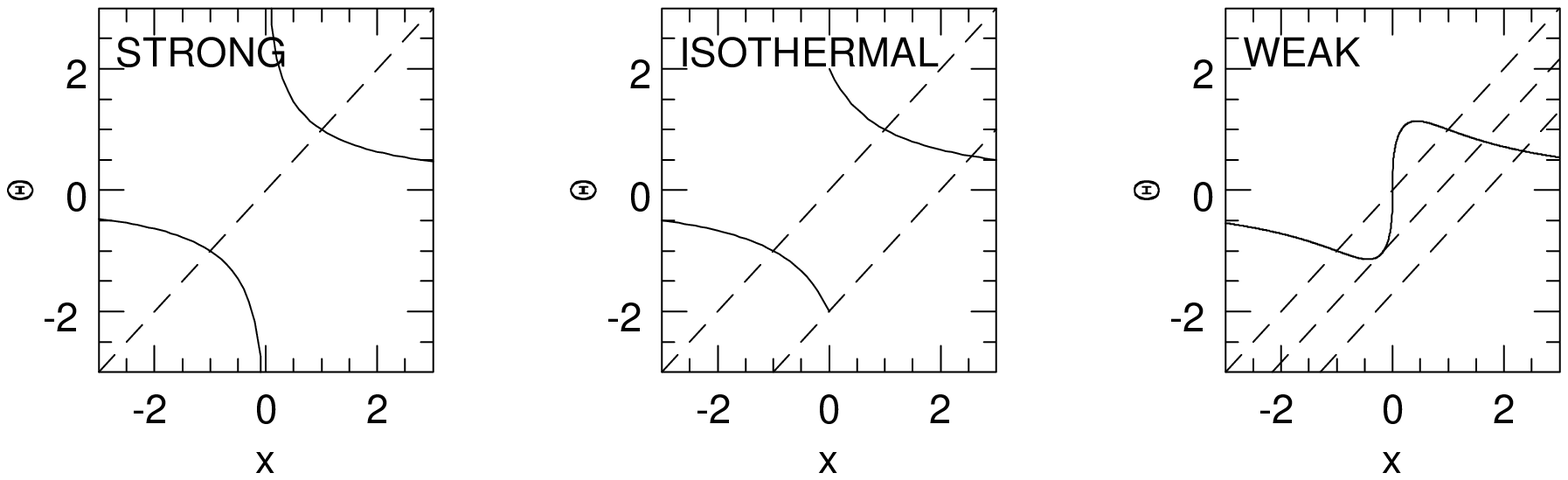,height=8.0truecm}}

\smallskip\noindent
\caption{{\bf Figure 1.} Multiple image diagrams for strong,
isothermal and weak cusps. Typical $x-y$ curves are shown in dashed
lines for a number of source positions. Multiple imaging takes place
for the source position $y_1$ if the line $x-y_1$ intersects the
deflection angle curve more than once. In the strong cusp, no matter
how distant the source position, the observer sees at least two
images. In the isothermal cusp, multiple imaging only takes place if
the source is sufficiently close to the cusp.  For sources far enough
away, only one image is observed. In the weak cusp, three images are
seen for source close to the cusp, one image for distant sources. The
change-over in the number of images occurs as the source crosses the 
radial caustic, as shown.}
\endfigure

This change in our understanding of the structure of galaxies and
clusters motivates our study of geometrically thin, centrally cusped
gravitational lenses. Hitherto, the emphasis in gravitational lensing
studies has largely been placed on non-singular models, such as
softened isothermal spheres and ellipsoids or Plummer models (e.g.,
Blandford \& Kochanek 1987; Kochanek 1996; Keeton \& Kochanek 1997;
Schneider, Ehlers \& Falco 1992, chap. 8). For such non-singular
lenses, it is well-known that the total number of images is odd and
that the number of even parity images exceeds the number of odd parity
images by one (e.g., MacKenzie 1985; Burke 1981; Schneider, Ehlers \&
Falco 1992, chap. 5). This paper begins with an outline of general
results on geometrically thin, centrally cusped lens models
appropriate for galaxies and clusters. In Section 3, a new and simple
family of axisymmetric cusped lenses, the double power-law lenses, is
presented, together with an application to radial arcs in clusters.
Section 4 studies one member of this family, the isothermal
double-power lens, in detail. It is worthy of special scrutiny, as the
lens equation is solvable for any source position. This property is
possessed by only two other known circularly symmetric lenses (the
Schwarzschild lens and the isothermal sphere). Section 5 examines a
family of density cusps of infinite extent, with equipotentials that
are similar concentric ellipses. The isothermal model in this family
has already been extensively investigated (e.g., Kassiola \& Kovner
1993). The remaining members have not been examined in detail as
isolated lenses, although the statistical properties of ensembles of
these models have been studied in two classic papers (Blandford \&
Kochanek 1987; Kochanek \& Blandford 1987).

In all of the following sections, we use the notation and conventions
of the admirable text-book of Schneider, Ehlers \& Falco (1992). In
particular, we use $(x_1,x_2)$ as Cartesian coordinates in the lens
plane and $(y_1,y_2)$ as Cartesian coordinates in the source
plane. The Poisson equation relating the convergence $\kappa$ and the
deflection potential $\psi$ is scaled to read $\nabla^2 \psi =
2\kappa$. The deflection angle $\Theta$ is just $\nabla \psi$.

\eqnumber =1
\def\chaphead{\hbox{2.}}
\section{General Theorems} 
 
To orient ourselves, we start with the simpler case of axisymmetric
lenses with density cusps. Theorems for general cusped lenses are
deduced in Section 2.2.

\subsection{Axisymmetric Lenses}

Let us consider a geometrically thin, axisymmetric lens with
convergence $\kappa(R)$, where $R = |{\u x}|$. For an axisymmetric
lens, the ray trace equation is one-dimensional, as all light rays
emanating from the source and received by the observer lie in the
common plane incorporating the observer, lens and source. Let us
choose coordinates in the lens plane so that ${\u x} = (x,0)$ and in
the source plane so that ${\u y} = (y,0)$. The surface density is
assumed to be piecewise continuous everywhere except at the central
cusp $R=0$.  We suppose that the deflection angle vanishes
asymptotically, or, equivalently, the convergence falls off at large
radii faster than $1/R$, i.e.,
$$\lim_{R \rightarrow \infty} R \kappa(R) =0.\eqno\new$$
Let the convergence diverge at the origin like
$$\lim_{R \rightarrow 0} \kappa(R) = O(R^{-\gamma}).\eqno\new$$
Then:

\noindent
(1) If $1< \gamma <2$, multiple imaging occurs for any position of the
source $y$, no matter how large. The model is said to have a strong
cusp.

\noindent
(2) If $\gamma =1$, the cusp is isothermal. Multiple imaging occurs for
some source positions. For sufficiently large values of $y$, there is
only one image. For sufficiently small values of $y$, there are at least
two images.

\noindent
(3) If $0< \gamma <1$, multiple imaging occurs for some source
positions.  For sufficiently large values of $y$, there is only one
image. For sufficiently small values of $y$, there are at least three
images. The model is said to have a weak cusp.

\noindent
(4) If $\gamma \le 0$, the model is uncusped in projection.  A
necessary and sufficient condition for multiple imaging is that the
central convergence $\kappa(0)$ must exceed unity (Schneider, Ehlers
\& Falco 1992, p. 236). We remark that models which are uncusped in
projection may still possess a mild cusp in the three-dimensional
luminosity density.

These statements are readily proved. As the cusp is approached,
the deflection angle $\Theta = \nabla \psi$ behaves like
$$\lim_{R \rightarrow 0} \Theta = O(R^{-\gamma+1}).\eqno\new$$
Near the cusp, the deflection angle vanishes if $\gamma <1$, tends to
a constant value if $\gamma =1$ and diverges if $\gamma >1$.  Now, the
deflection angle is an odd function of $x$. So, this implies that
$\Theta$ is discontinuous at $x=0$ for isothermal cusps, while
$\Theta$ possesses a singularity at $x=0$ for strong cusps.  The
gradient of the deflection angle behaves like
$$\lim_{R \rightarrow 0} {\d \Theta \over \d R} = O(R^{-\gamma}).
\eqno\new$$
This diverges for all cusped models.

Figure 1 shows multiple image diagrams for strong, isothermal and weak
cusps in typical centrally condensed galaxy or cluster models. The
graphs show the deflection angle $\Theta$, together with $x-y$ lines
for a number of values of the source position. The location of the
images is given by the solutions of the lens equation, and therefore
by the intersection of the $x-y$ lines with the deflection angle. So,
for strong cusps, it is evident from Fig.~1 that there are always two
images, irrespective of the distance of the source. Isothermal cusps
have the curious property that the number of images can change by one,
as the source position is moved outward across a curve that we shall
call {\it the pseudo-caustic}. This is illustrated in the second panel
of Fig.~1.  This curiosity has previously been noticed in analyses of
the singular isothermal sphere and ellipsoid (Kovner 1987a; Kormann,
Schneider \& Bartelmann 1994; Schneider, Ehlers \& Falco 1992,
p. 243).  Weak cusps always exhibit multiple imaging for sufficiently
small $y$. As the source is moved away, there comes a critical point
when it crosses the radial caustic and the number of images diminishes
by two. This is shown in the third panel of Fig.~1.

Let us note that the observations of the centres of early-type
galaxies indicate that strong, weak and isothermal cusps all can
occur. For example, table 2 of Faber et al. (1997) gives the cusp
slope $\gamma$ for a sample of 61 elliptical galaxies. The
observationally fitted values of $\gamma$ lie in the range $0.0 <
\gamma < 1.21$. NGC 1199 is an example of a galaxy with a strong cusp,
while NGC 4467 has an isothermal cusp. Weak cusps are the most
predominant in Faber et al.'s (1997) sample. The cosmological
simulations of Navarro, Frenk \& White (1996) suggest that dark matter
haloes possess a universal density law with a logarithmically singular
surface density. Subsequently, numerical investigations (Fukushige \&
Makino 1997; Moore, Governato, Quinn, Stadel \& Lake 1997) together
with theoretical arguments (Evans \& Collett 1997) indicated that the
original work had underestimated the severity of the cusp. The weight
of the evidence now suggests that dark haloes have three-dimensional
density cusps like $\rho \sim r^{-4/3}$. In projection, this becomes
$\kappa \sim r^{-2/3}$ and so lies within the weak cusp r\'egime.

\beginfigure{2}

\centerline{\psfig{figure=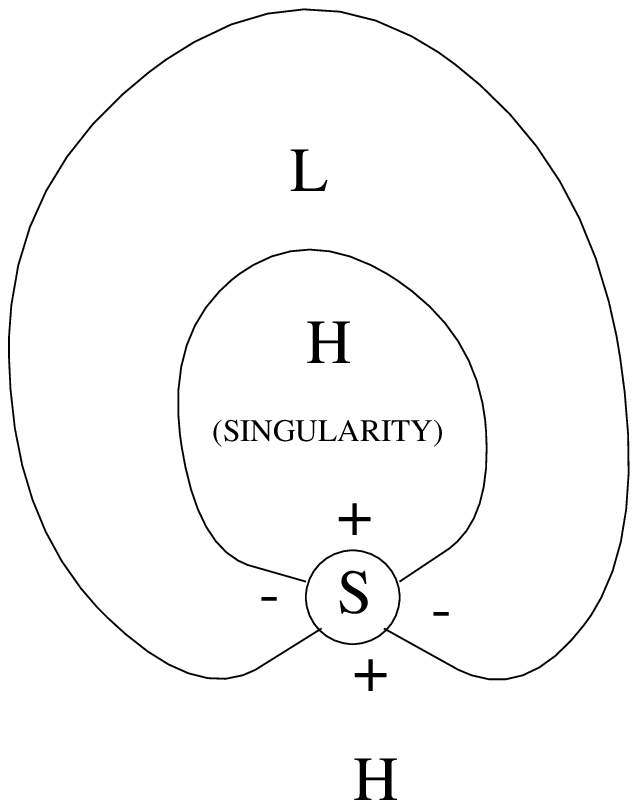,height=0.5\hssize}}

\smallskip\noindent
\caption{{\bf Figure 2.} The only possible two image topography (the
limacon) of the Fermat surface for lenses with strong density
cusps. The saddle point is marked by S and the $+$ and $-$ signs
indicate regions where the Fermat surface is higher (H) and lower (L)
than the saddle. An image of type II occurs at the saddle, an image of
type I at the minimum. The density singularity is marked.}

\endfigure

\begintable*{1}%
\caption{{\bf Table 1.} Some simple potentials corresponding to
the convergence of the double power-law family as given in eqs.~(3.3)
and (3.8). (Below, $\ell$ and $n$ are natural numbers. The summands
are empty if $\ell=0$ or $n=0$.)}
\halign{#\hfil&\quad#\hfil&\quad#\hfil\cr
\noalign{\hrule}
\noalign{\vskip0.3truecm}
$L=1$         &$N=2n$  
              &$\psi(R)=\displaystyle {8C\alpha^2\over 2n+1}
               \Bigl[\log(1+\sqrt{R^{1/\alpha}+1}) - 
               \sum_{k=1}^n {1\over 2k-1}
               {1\over ( 1+ R^{1/\alpha})^{k-1/2}}\Bigr]$  \cr
\noalign{\vskip0.2truecm}
$L=1$         &$N=2n+1$
              &$\psi(R)=\displaystyle {2C\alpha^2 \over n+1}
              \Bigl[\log(1+R^{1/\alpha}) - 
              \sum_{k=1}^n {1\over k}
              {1\over ( 1+ R^{1/\alpha})^k}\Bigr]$  \cr
\noalign{\vskip0.2truecm}
$L=2\ell$       &$N=1$
                &$\psi(R) =\displaystyle {8C\alpha^2\over 2\ell+1}
               \Bigl[\log(R^{1/(2\alpha)}+\sqrt{R^{1/\alpha}+1}) - 
               \sum_{k=1}^\ell {1\over 2k-1}
               \Bigl({R^{1/\alpha} \over 1+ R^{1/\alpha}}\Bigr)^{k-1/2}
               \Bigr]$  \cr
\noalign{\vskip0.2truecm}
$L=2\ell +1$  &$N=1$
              &$\psi(R) = \displaystyle {2C\alpha^2 \over \ell+1}
             \Bigl[\log(1+R^{1/\alpha}) - 
             \sum_{k=1}^\ell {1\over k}
             \Bigl({R^{1/\alpha} \over 1+ R^{1/\alpha}}\Bigr)^k\Bigr]$ \cr
\noalign{\vskip0.1truecm}
\noalign{\hrule}
}
\endtable

\subsection{General Lenses and Index Theorems}

There is a general theorem for non-singular lenses that states that
the total number of images is odd and that the number of even parity
images exceeds the number of odd parity images by one (MacKenzie 1985;
Burke 1981; Fukuyama \& Okamura 1997).  The image type is classified
by studying the Jacobian matrix of the Fermat potential $\phi$. If
both the eigenvalues are positive or both are negative, then the
images are of type I and III respectively.  Such images have even
parity and correspond to minima or maxima of the Fermat potential.  If
one of the eigenvalues is positive and one negative, the image is of
type II. Such images have odd parity and correspond to saddle-points
of the Fermat potential. The odd number theorem is usually stated in
the form
\eqnam{\oddno}
$$\ni - \nii + \niii = 1, \eqno\new$$
where $\ni$ is the number of images of type I and so on. It follows
from this that the total number of images ($\ni + \nii + \niii$) is
necessarily odd. 

To set the notation, let us quickly re-derive the odd number theorem
for non-singular lenses. Introducing $z = x_1 + i x_2$, $\zbar = x_1 -
i x_2$, then the lens equation becomes just (e.g., Witt 1990)
$${\partial \phi \over \partial \zbar} = \phi_\zbar = 0.\eqno\new$$
Let $\cz$ be a contour enclosing some point P in the complex $z$
plane. The index of the map $\omega = \phi_\zbar(z)$ at P is
$$\Ind (\phi_\zbar, P) = {1\over 2 \pi i} \oint_{\cw} d \ln \omega,
\eqno\new$$
where $\cw$ is the image of $\cz$. So, the index is the number of
times the image of P in the complex $\omega$-plane is encircled, as P
in the complex $z$-plane is encircled once in the anti-clockwise
direction. If $\cz$ does not enclose any zeroes or singular points,
the index vanishes. It is straightforward to show (e.g., Schneider,
Ehlers \& Falco 1992, p. 174) that the index of an extremum of the
Fermat potential is $+1$, whereas the index of a saddle point is $-1$.
Provided the deflection angle tends to zero at large radii, then it
must be true asymptotically that
$$\phi_\zbar = \fr12 | z - \zs | + o(|z|),\eqno\new$$
where $\zs = y_1 + i y_2$ is the complex position of the source.  So,
the point at infinity is a pole and it has an index of unity. By
shrinking the contour so that it becomes enmeshed with the singular
points and zeroes, it follows that the sum of all the indices must be
equal to unity. If the lens is non-singular, there are no poles and
only the contributions of the zeroes must be taken into account, i.e.,
$$1 = \Ind (\phi_\zbar, \infty) = \sum_i  Ind (\phi_\zbar, P_i)
= \ni - \nii + \niii.\eqno\new$$
This is the odd number theorem and holds good only for non-singular 
lenses with convergence falling off faster than $1/|z|$ at large
radii.

What is the analogue of this theorem for general lenses with density
cusps? A careful examination of the density singularity at the origin of
the z-plane is required to establish its index. Near the origin, the
lens equation can always be cast into the form
\eqnam{\cuspato}
$$\phi_\zbar = \fr12 | z - \zs | + {A z^{1-\gamma/2}\over 
\zbar^{\gamma/2}},\eqno\new$$
where $A$ is a constant (for circular lenses) or a real function 
of the phase (for non-axisymmetric lenses) and the sub-dominant
terms have been neglected.  Substituting $z = \epsilon \exp (i
\theta)$, this becomes
\eqnam{\epsy}
$$\phi_\zbar = \fr12 | \epsilon \exp(i\theta) - \zs | + A \epsilon^{1 -\gamma}
\exp(i\theta).\eqno\new$$
Letting $\epsilon \rightarrow 0$ so that the contour is wrapped
tightly around the origin, we deduce that the first term is dominant
in the case of weak cusps ($0< \gamma <1$) and the second term is
dominant in the case of strong cusps ($1< \gamma <2$).  In the limit,
the vector field $\phi_\zbar$ is constant for weak cusps, so the the
index of the origin is $0$ and the odd number theorem (\oddno) still
holds good. For strong cusps, a different situation obtains. The index
of the origin is $+1$, as the vector field is radial in the limit
$\epsilon \rightarrow 0$. Lens models with strong cusps have an even
number of images, as
\eqnam{\evenno}
$$ \ni - \nii + \niii = 0.\eqno\new$$
The number of even parity images is equal to the number of odd
parity images.

The instance of isothermal cusps ($\gamma =1$) is more subtle. Now
both terms in (\cuspato) matter and the index of the origin depends on
the position of the source $\zs$. As $\epsilon \rightarrow 0$, the
lens equation (\epsy) becomes
\eqnam{\psudoc}
$$\phi_\zbar = \fr12 | \zs | + A \exp(i\theta),\eqno\new$$
where $A = \psi_\zbar$ which depends only on the phase for models with
isothermal cusps.  If $| \zs | < 2|\psi_\zbar|$, then the index of the
origin is $+1$, otherwise the index of the origin is $0$. This
equation therefore defines a curve in the source plane -- the
pseudo-caustic. If the source is outside the pseudo-caustic, there is
an odd number of images according to (\oddno). Crossing the
pseudo-caustic produces an additional, single image of type II
initially at the origin. Once the source is within the pseudo-caustic,
there is an even number of images according to (\evenno). Note that
the pseudo-caustic differs from a true caustic because the
magnification of a point source is finite (not infinite) and because
the number of images changes by one (not two).  The analogues in the
lens plane are the pseudo-critical curves.

As is well-known (Blandford \& Narayan 1986; Schneider, Ehlers \&
Falco 1992, p. 178), the topography of the Fermat surface for
non-singular lenses can be classified in terms of the local structure
at saddle points.  For three image geometries, the shape of the
critical isochrone corresponds either to a lemniscate or a
limacon. For five image geometries, there are six possible
topographies given by joining lemniscates and limacons. These theorems
still hold for lenses with weak density cusps. But, if the lens
possesses a strong density cusp, then the topography of the Fermat
surface is more restricted. For two image geometries, the only
possibility is the limacon illustrated in Fig.~2. One of the images
occurs at the saddle point marked by S, one in the basin of the Fermat
surface. The density cusp corresponds to the highest point. For four
image geometries, there are five possible topographies, given by
joining at least one limacon to either a lemniscate or a limacon.

\beginfigure{3}

\psfig{figure=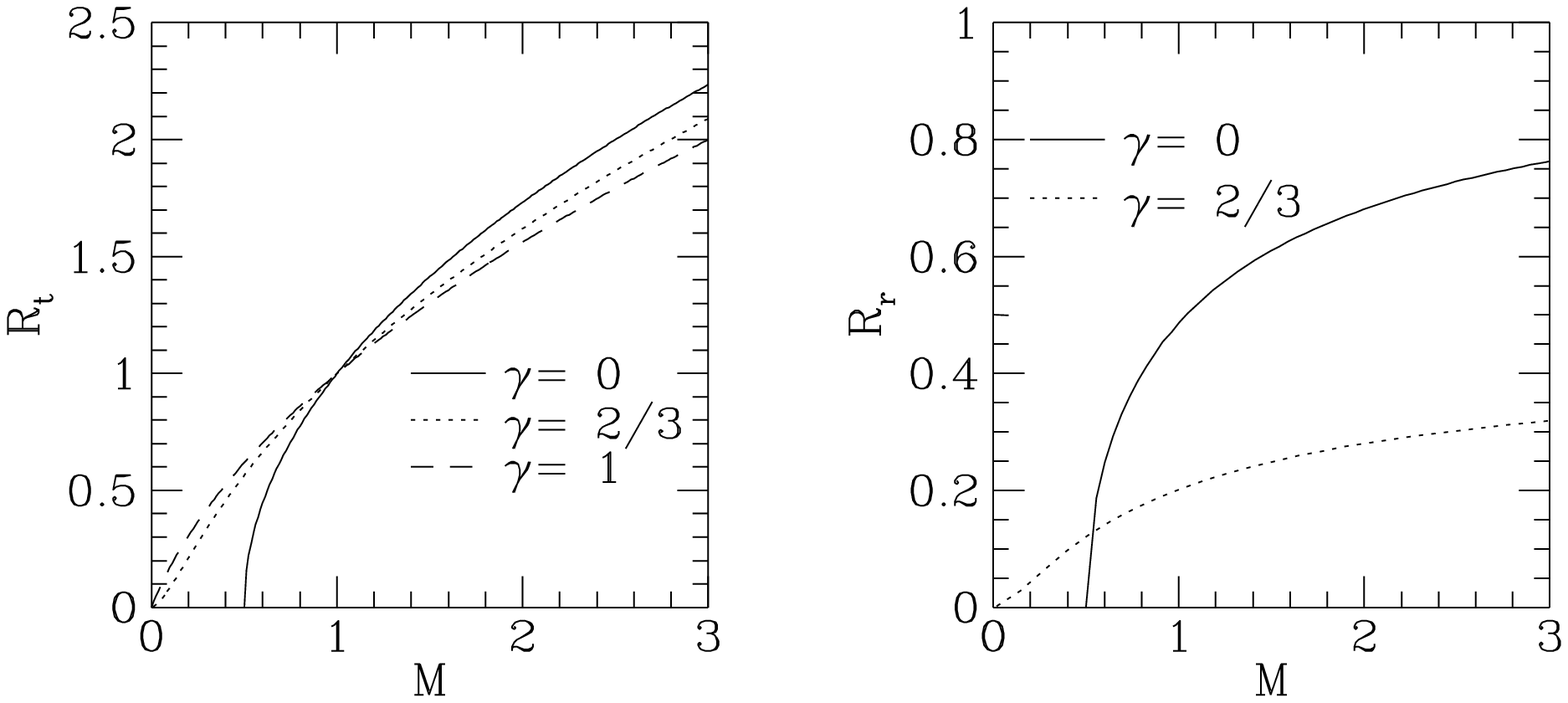,width=\hssize}

\smallskip\noindent
\caption{{\bf Figure 3.} The variation of the radii of the tangential
and radial critical circles with total projected mass for various
values of $\gamma$. Note that radial critical curves and caustics do not
exist if $\gamma \ge 1$, as is evident on referring back to the multiple
imaging diagrams of Fig.~1.}

\endfigure

\eqnumber =1
\def\chaphead{\hbox{3.}}
\section{The Double Power-Law Lenses}

We now present a new family of axisymmetric lenses with central
density cusps, the double power-law lenses. Section 3.1 explains our
strategy for locating the models with simple potential-convergence
pairs. The following sections examine properties and applications of
the models in detail.

\subsection{The Potential-Convergence Pair}

Let us consider the family of cusped lenses with surface density $\Sigma$
as a function of projected radius $\xi$ given by
\eqnam{\sdensity}
$$\Sigma (\xi) = {C \xic^\beta \over \xi^\gamma (\xi^{1/\alpha}
+\xic^{1/\alpha})^{\alpha(\beta-\gamma)}}.
\eqno\new$$
Here, ($\alpha,\beta,\gamma$) are positive constants and $\xic$ is a
scale-length. By choosing the scale in the lens plane as $\xic$, then
the convergence $\kappa$ is defined as (e.g., Schneider, Ehlers \&
Falco 1992, chap. 5)
$$\kappa(R) = {\Sigma (R\xic)\over \Sigmac}, \qquad \Sigmac = {c^2\ds
\over 4 \pi G \dd \dds},\eqno\new$$
where $\Sigmac$ is the critical surface density and $\ds, \dd$ and
$\dds$ are the observer-source, the observer-lens and the lens-source
distances respectively.  For our model (\sdensity), this gives
\eqnam{\convergence}
$$\kappa (R) = {C\over R^\gamma (R^{1/\alpha} +1)^{\alpha(\beta-\gamma)}}.
\eqno\new$$
At small radii, the convergence is cusped like $R^{-\gamma}$, whereas
in the outer parts, it falls like $R^{-\beta}$. The parameter $\alpha$
controls the extent of the transition region between the cusp and the
envelope. The lens models (\convergence) are clearly inspired by the
`Nuker' profile introduced by Gebhardt et al. (1997) and Faber et
al. (1997).  These authors show that it provides a compact description
of the surface brightness of early-type galaxies.  The lens models
(\convergence) are also related to the three-dimensional elliptical
galaxy models introduced by Hernquist (1990) and studied by Zhao
(1996). The normalisation constant $C$ is related to the total mass
$\mtot$ by
$$\mtot = 2 C \pi \xic^2 \Sigmac \alpha
B(\alpha[2-\gamma],\alpha[\beta-2]),\eqno\new$$
where $B(x,y) = \Gamma(x) \Gamma(y) / \Gamma (x +y)$ is the complete
Beta function.  For convenience, we also define a dimensionless mass
$M$ such that
$$M = {\mtot \over 2 \pi \xic^2 \Sigmac}.\eqno\new$$
For the total mass to converge, we must insist that $\beta >2$ and
that $\gamma <2$.

The solution of the two-dimensional Poisson equation is familiar from
the theory of the Newtonian potential of circular cylinders as (e.g., 
Routh 1892; Ramsey 1940)
\eqnam{\newtonian}
$$\psi(R) = 2\log R \int_0^R \d r\, r \kappa(r) + 2\int_R^\infty \d r
            \,r \kappa(r)\log r.\eqno\new$$ 
Of course, the first term is the potential of the circular cylinders
interior to $R$, which attract as if their mass were concentrated on
axis. The second term is the potential of the circular cylinders
exterior to $R$. Integrating (\newtonian) by parts, we obtain:
\eqnam{\generalpotential}
$$\eqalign{\psi(R) =& 2\log R \int_0^\infty \d r\,r \kappa(r) \cr
            &\qquad\qquad\qquad + 2\int_R^\infty {\d r\over r} 
            \int_r^\infty \d r'\,r' \kappa(r').\cr}\eqno\new$$ 
Substituting the convergence (\convergence), we see that in general
the potential is not elementary. Of course, there is an enormous
advantage to working with models with at least simple deflection
angles and preferably simple potentials.  So, let us first outline
conditions under which these quantities are reducible to a finite
number of elementary functions. It is helpful to introduce two
constants $L$ and $N$ such that
\eqnam{\integers}
$$L = 2\alpha[2-\gamma] -1, \qquad N = 2\alpha[\beta-2] -1.\eqno\new$$
Making the substitution $R = \tan^{2\alpha} \theta$, then the 
deflection angle is
\eqnam{\deflection}
$$\eqalign{\Theta =& {\d \psi\over \d R} 
                  = {4C\alpha\over R}\int_0^\theta \d\theta'
                   \sin^L \theta'\cos^N \theta'\cr
                  =& {2C\alpha\over R} B( L/2 + 1/2, N/2 + 1/2; 
                   \sin^2\theta).\cr}\eqno\new$$
Here, $B(x,y;z)$ is the incomplete Beta function (see e.g., Abramowitz 
\& Stegun 1965; Gradshteyn \& Rhyzik 1980). Amongst other instances, it
reduces to a finite numer of elementary functions if either (1) $L$ is
an odd natural number and $N$ is arbitrary, or vice versa, or (2) $L$
is an even natural number (or zero) and $N$ is any natural number (or
zero), or vice versa. 

The potential can similarly be reduced to
\eqnam{\simplerpot}
$$\eqalign{\psi =& 4C\alpha \log R \int_0^\pibytwo \d \theta'
                   \sin^L \theta' \cos^N \theta' \cr
            +& 8C\alpha^2\int_\theta^\pibytwo {\d \theta'\over
            \sin \theta' \cos \theta'}\int_{\theta'}^\pibytwo
            \d \phi \sin^L \phi \cos^N \phi.\cr}\eqno\new$$ 
A consultation of the discussion in sections 2.51 and 2.52 of Gradshteyn
\& Rhyzik (1980) reveals that the indefinite integration in (\simplerpot) 
is elementary if $L$ and $N$ are natural numbers (or zero), at least
one of which is odd.  Let us note explicitly that if $L$ and $N$ are
both even natural numbers (or zero), the potential is not elementary
as it ultimately depends on the transcendent integral (2.644.5) of
Gradshteyn \& Rhyzik (1980).

Some simple examples of lens models with convergence (\convergence)
possessing elementary potentials are given in Table 1. Henceforth, we
specialise to the case $L=1, N=1$ for definiteness.

\subsection{The $L=1, N=1$ Family}

Here, the potential-convergence pair is just
\eqnam{\neatfamily}
$$\kappa = {M (2- \gamma)\over R^{\gamma}(1+ R^{2-\gamma})^2},\qquad
\psi = {2M\over 2-\gamma} \log ( 1 + R^{2-\gamma}).\eqno\new$$
If $1 < \gamma < 2$, the lens has a strong cusp.  This becomes
isothermal when $\gamma = 1$.  For $0 < \gamma < 1$, the lens
possesses a weak cusp.  The instance $\gamma = 0$ is the uncusped
Plummer model, discussed in detail in Schneider, Ehlers \& Falco
(1992, chap. 8).  Models with $\gamma < 0$ are unrealistic, as the
density is not centrally concentrated.

The scaled deflection angle $\Theta$ at a displacement $x$ in the lens
plane is
$$\Theta = \nabla \psi = {2M\sign (x) |x|^{1 - \gamma}\over 
                          1 + |x|^{2-\gamma}},\eqno\new$$
where $|x| =R$. In the lens plane, the radius of the tangential
critical circle $\rt$ is given by
\eqnam{\tcc}
$$\rt^2 (1+ \rt^{2-\gamma}) = 2M\rt^{2-\gamma}.\eqno\new$$
Generally, this implicit equation cannot be solved analytically.
It is exactly solvable when $\gamma = 1$ to give
$$\rt = \fr12 \sqrt{1 + 8M} -\fr12,\eqno\new$$
when $\gamma = 0$ to give
$$\rt = \sqrt{2M-1}.\eqno\new$$
The tangential critical circles are just the radii of the bright
Einstein rings formed when the source is exactly coincident with
the lens. The radial critical curves have radius $\rr$ given by
\eqnam{\rcc}
$$(1 + \rr^{2-\gamma})^2 = 2M((1-\gamma)\rr^{-\gamma} -
                                  \rr^{2 - 2\gamma}).\eqno\new$$
When $\gamma =0$, this is solvable to give
$$\rr^2 = \sqrt{M (4 + M)} - M -1 .\eqno\new$$
When $\gamma \ge 1$, this equation has no positive root and there is
no radial critical curve. As is evident from the multiple imaging
diagrams of Fig.~1, isothermal and strong cusps do not have radial
critical curves.  Figure 3 shows the variation of the critical curves
with total projected mass. The curves are labelled with the value of
$\gamma$. The graph shows that as $M \rightarrow \infty$, the
tangential critical curves always tend to $\sqrt{2 M}$, which is the
Einstein ring radius of a point mass (let us recall $M$ is mass
measured in units of the $2\pi \xic^2 \Sigmac$).

\beginfigure*{4}

\psfig{figure=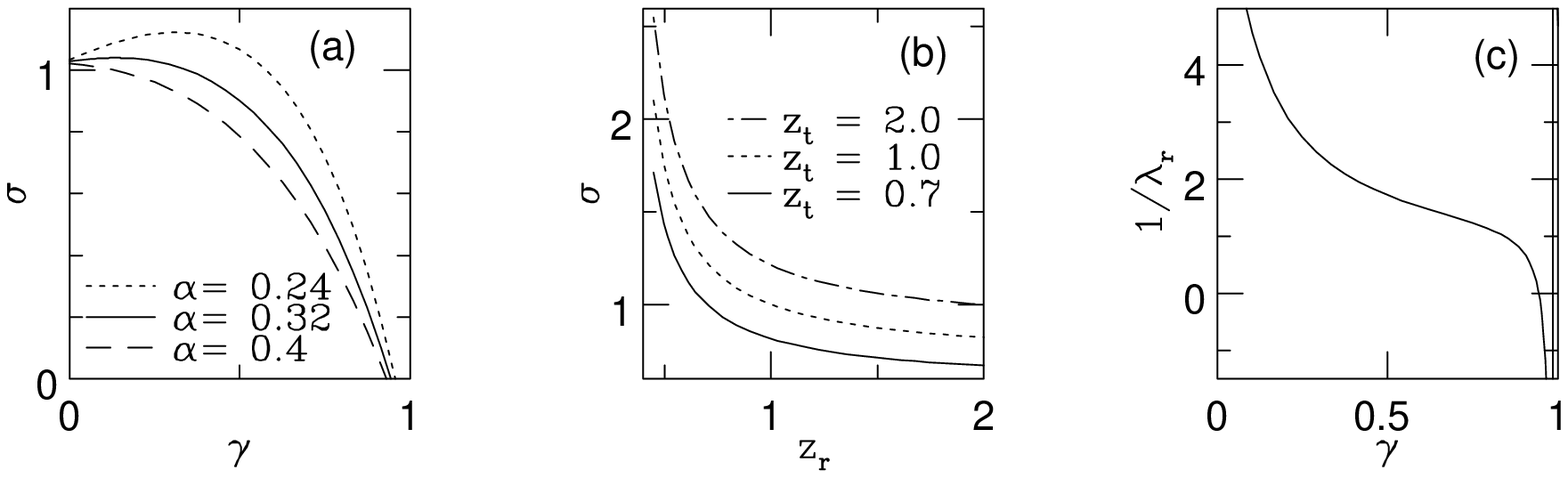,width=\hdsize}

\smallskip\noindent
\caption{{\bf Figure 4.} (a) $\sigma$ versus $\gamma$ for different
values of $\alpha$. (b) $\sigma$ versus radial source redshift $z_{\rm r}$
for various values of tangential source redshift $z_{\rm t}$. (c) Radial
magnification at the location of the tangential arc plotted versus $\gamma$
for $\rt = 0.2$ and $\rr = 0.0645$.}

\endfigure

The tangential caustic coincides with the origin in the source plane
$y_t=0$. The radial caustic exists only for weak cusps and is
$$y_r = { \rr ( \gamma + 2\rr^{2-\gamma}) \over
             \gamma -1 + \rr^{2-\gamma}}.\eqno\new$$
The radial caustics separates the source domains ($|y|
< \yr $) generating three images from those ($|y| > \yr$) generating
just one. The magnification is 
$$\eqalign{\mu=&\Bigl[ 1 - {2M R^{-\gamma}\over 1 + R^{2-\gamma}}
                \Bigr]^{-1} \times \cr
    &\Bigl[ 1 - {2M((1-\gamma)R^{-\gamma} -
     R^{2-2\gamma})\over (1 + R^{2-\gamma})^2}
     \Bigr]^{-1}.\cr}\eqno\new$$ 
When $|y| < \yr$, the image at $x >\rt$ is of type I, the image at
$-\rt<x<-\rr$ is of type II and the image at $-\rr < x<0$ is of type
III. The two components of the shear are
$$\eqalign{\gamma_1 =& {M (x_2^2 - x_1^2)(\gamma + 2 R^{2-\gamma})
                        \over R^{\gamma} (1+R^{2-\gamma})^2},\cr
           \gamma_2 =& {-2M x_1x_2(\gamma + 2R^{2-\gamma})
                        \over R^{2+\gamma} (1 + R^{2-\gamma})^2}.\cr}
                        \eqno\new.$$

\subsection{An Application: Radial Arcs}

Radial arcs are images of galaxies distorted by foreground
clusters. They are elongated in the radial direction, in distinction
to the more commonly occurring tangential arcs.  Bartelmann (1996)
described how, in a situation where both radial and tangential arcs
are observed, the positions of the arcs can be used to determine the
parameters of the model potential representing the lens. Bartelmann
used the three-dimensional density profile suggested by the
cosmological simulations of Navarro, Frenk \& White (1996), namely:
\eqnam{\bartden}
$$\rho(r) = {\rhos \rc^3\over r(\rc + r)^2},\eqno\new$$
where $r$ is the three-dimensional radius and $\rhos$ and $\rc$ are
density and length scales. He projected this to obtain the surface
density and hence the lens model. He showed that if both the positions
and redshifts are available for the tangential and radial arcs, then
it is possible to find values for the density and length scales. If
the redshifts of the arcs are not known, it is possible to constrain
them.  In particular, Bartelmann (1996) pointed out a problem with the
density profile (\bartden) -- for tangential arcs lying within the
scale radius, the lens model predicts large values for the radial
magnification, which in turn implies that the corresponding sources
must be surprisingly thin in the radial direction.

Can the more general family of surface densities (\sdensity) alleviate
the problem described by Bartelmann? In addition to the arbitrary
density and length scales (in our notation $M$ and $\xic$), we have
the extra parameter $\gamma$ describing the cusp. By varying the value
of $\gamma$, can we avoid the problem of large radial magnification at
the tangential arcs?  Equations (\tcc) and (\rcc) give the positions
of the radial and tangential critical curves for the family of lenses
with convergence (\neatfamily). Assuming that the sources producing
the radial and tangential arcs are at different redshifts, we rewrite
(\tcc) and (\rcc) as
\eqnam{\rccn}
$$\eqalign{\rt^2 (1+ \rt^{2-\gamma}) =& 2M_{\rm t}\rt^{2-\gamma},\cr
(1 + \rr^{2-\gamma})^2 =& 2M_{\rm r}((1-\gamma)\rr^{-\gamma} -
                                  \rr^{2 - 2\gamma}),\cr}\eqno\new$$
where the subscripts on $M$ indicate its redshift dependence. Defining
the ratio $\sigma$ by
\eqnam{\siga}
$$\sigma \equiv {\Sigma_{\rm cr,r}\over \Sigma_{\rm cr,t}} = 
{M_{\rm t}\over M_{\rm r}},\eqno\new$$
we obtain, using (\rccn),
\eqnam{\sigb}
$$\sigma(\alpha) =
{(1+\rt^{2-\gamma})(1-\gamma-\rt^{2-\gamma}\alpha^{2-\gamma})\over 
\alpha^{\gamma}(1+\alpha^{2-\gamma}\rt^{2-\gamma})^2},\eqno\new$$
where $\alpha = \rr/\rt$. The value of $\alpha$ is known from the
positions of the radial and tangential arcs. If the redshifts of the
arcs are also available, then we can use (\siga) to find the
corresponding value of $\sigma$. Choosing a value for $\gamma$, we can
solve (\sigb) for $\rt$. Since $\rt = \xi_{\rm t}/\xic$, and we can
observe the value of the physical position of the tangential arc
$\xi_{\rm t}$, we can obtain the scale $\xic$. Finally, (\rccn) can be
used to obtain the value of the density scale.

Often the redshifts are unavailable and we must proceed more
indirectly. As an example, let us consider the lensing cluster MS
2137, which is at a redshift of $z_c = 0.315$. In this cluster, a
tangential arc has been observed at $\xi_{\rm t} = 15''.5$ from the
cluster centre. Fort et al. (1992) detected a radial arc in the same
cluster at $\xi_{\rm r} = 5''.0$ from the cluster centre. To proceed,
we assume that $\xic \sim 250 h^{-1}\kpc$, a value obtained from
numerical simulations of dark matter halos within the CDM cosmogony
(cf. Bartelmann 1996).  From the observed values of $\xi_{\rm t}$ and
$\xi_{\rm r}$, we conclude that $\alpha = 0.32$ and we find that $R_t
\sim 0.2$. In Fig.~4(a), we have plotted $\sigma$ as a function of 
$\gamma$ for $\rt = 0.2$ with $\alpha = 0.32$ (and also for
$0.75\alpha$ and $1.25\alpha$ to give a feel for the likely
uncertainties). By varying the value of $\gamma$, we can obtain a wide
range in $\sigma$. In Fig.~4 (b), we show $\sigma$ versus the redshift
$\zr$ of the radial arc for a number of different values of the
redshift $\zt$ of the tangential arc. This plot shows that if $\sigma
\lta 0.6$, the redshifts of both sources must be unreasonably large. 
We should therefore choose a value of $\gamma$ for which $\sigma
\sim 1$.

Another constraint on our choice of $\gamma$ is the radial
magnification produced at the position of the tangential arc. Assuming
the same value for $\xic$ as before, we find that $\rr =
0.0645$. Substituting this into (\rccn) we can find $M_{\rm r}$
as a function of $\gamma$ only. The value of the radial eigenvalue as
a function of position is then given by
\eqnam{\lambdar}
$$\lamr(\rr) = 1 -  {2M_{\rm r}(\gamma)((1-\gamma)\rr^{-\gamma} -
               \rr^{2 - 2\gamma})\over (1 + \rr^{2-\gamma})^2}.\eqno\new$$
The radial magnification is just $1/\lamr$. In Fig.~4 (c), the radial
magnification at the position of the tangential arc is plotted as a
function of $\gamma$. In order to obtain a value of the magnification
which is of order unity, we must choose $\gamma$ in the range $0.6 \lta
\gamma \lta 0.95$. Fig.~4 (a) shows that values of $\gamma$ in the range
$0.6 \lta \gamma \lta 0.7$ correspond to acceptable values of
$\sigma$. In other words, the problem pointed out by Bartelmann is
assuaged by a choice of a cluster lens that is more singular than the
dark halo profile of Navarro, Frenk \& White (1996).

\beginfigure{5}

\psfig{figure=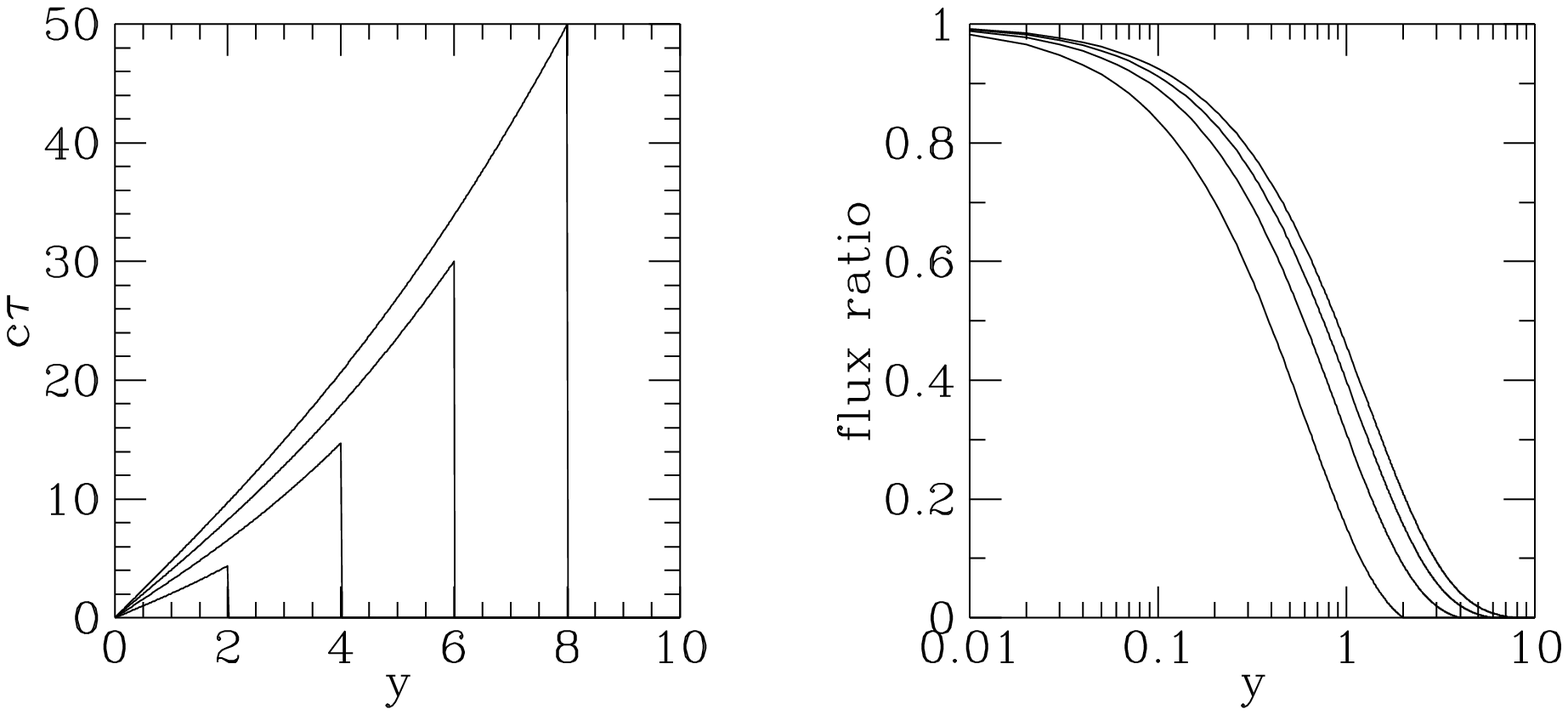,width=\hssize}

\smallskip\noindent
\caption{{\bf Figure 5.} The variation of the time delay $\tau$ 
and flux ratio of the images with separation of the source $y$ from
exact alignment with the isothermal double power-law lens. The curves
are shown for total mass $M = 1,2,3$ and $4$ respectively. As $y
\rightarrow 2M$, the second image becomes fainter and fainter. It
finally disappears when the source crosses the pseudo-caustic at $y =
2M$.  At this point, the notion of a time delay loses its
meaning. (Units used with $\xic^2
\ds(1 + \zd)/(c \dd \dds) =1$.)}

\endfigure

\beginfigure{6}

\psfig{figure=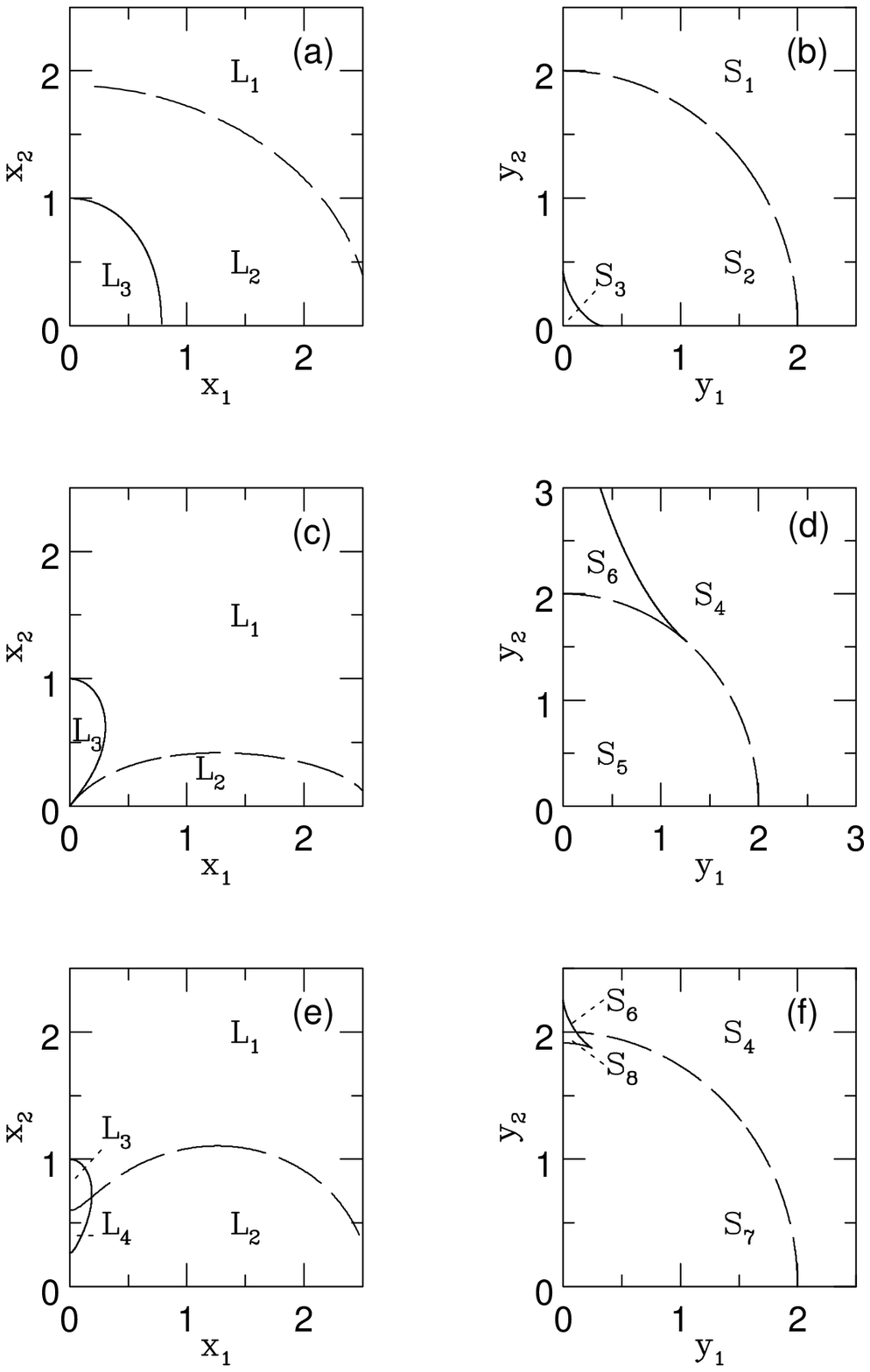,width=\hssize}

\smallskip\noindent
\caption{{\bf Figure 6.} The critical curve and caustic topologies
for the isothermal double power-law model perturbed with external
shear. In each case, the lens plane is shown on the left, the source
plane on the right. In panels (a) and (b), $C_1 > C_2 > 0$ and there
is one C point on each axis; (c) and (d), $C_1 >0, 0> C_2 >-0.5$ and
there is one C point on the $x_2$-axis and one S point at the origin;
in (e) and (f) $C_1 >0, C_2 < -0.5$ and there is a C point and an F
point on the $x_2$-axis.  The caustics and critical curves are drawn
in full lines, the pseudo-caustics and pseudo-critical curves are
drawn in broken lines.  (Units used with $M=1$.)}

\endfigure

\begintable{2}%
\caption{{\bf Table 2.} Image Positions as a Function of Source
Positions as Labelled on Figs~6 and 7. Roman numerals are used to
indicate the quadrant in which the image lies.}
\halign{#\hfil&\quad#\hfil\cr
\noalign{\hrule}
\noalign{\vskip0.3truecm}
S$_1$         &L$_1\rone$  \cr
\noalign{\vskip0.2truecm}
S$_2$         &L$_2\rone$, L$_3\rthree$  \cr
\noalign{\vskip0.2truecm}
S$_3$         &L$_2\rone$, L$_2\rtwo$,L$_3\rtwo$,L$_3\rthree$  \cr
\noalign{\vskip0.2truecm}
S$_4$         &L$_1\rfour$ \cr
\noalign{\vskip0.2truecm}
S$_5$         &L$_2\rthree$, L$_2\rfour$ \cr
\noalign{\vskip0.2truecm}
S$_6$         &L$_1\rthree$, L$_1\rfour$, L$_3\rthree$ \cr
\noalign{\vskip0.2truecm}
S$_7$         &L$_2\rfour$, L$_2\rthree$ \cr
\noalign{\vskip0.2truecm}
S$_8$         &2L$_2\rthree$, L$_2\rfour$, L$_4\rthree$ \cr
\noalign{\vskip0.2truecm}
S$_9$         &L$_1\rthree$ \cr
\noalign{\vskip0.2truecm}
S$_{10}$      &L$_2\rthree$, L$_3\rthree$ \cr
\noalign{\vskip0.1truecm}
\noalign{\hrule}
}
\endtable

\beginfigure{7}

\psfig{figure=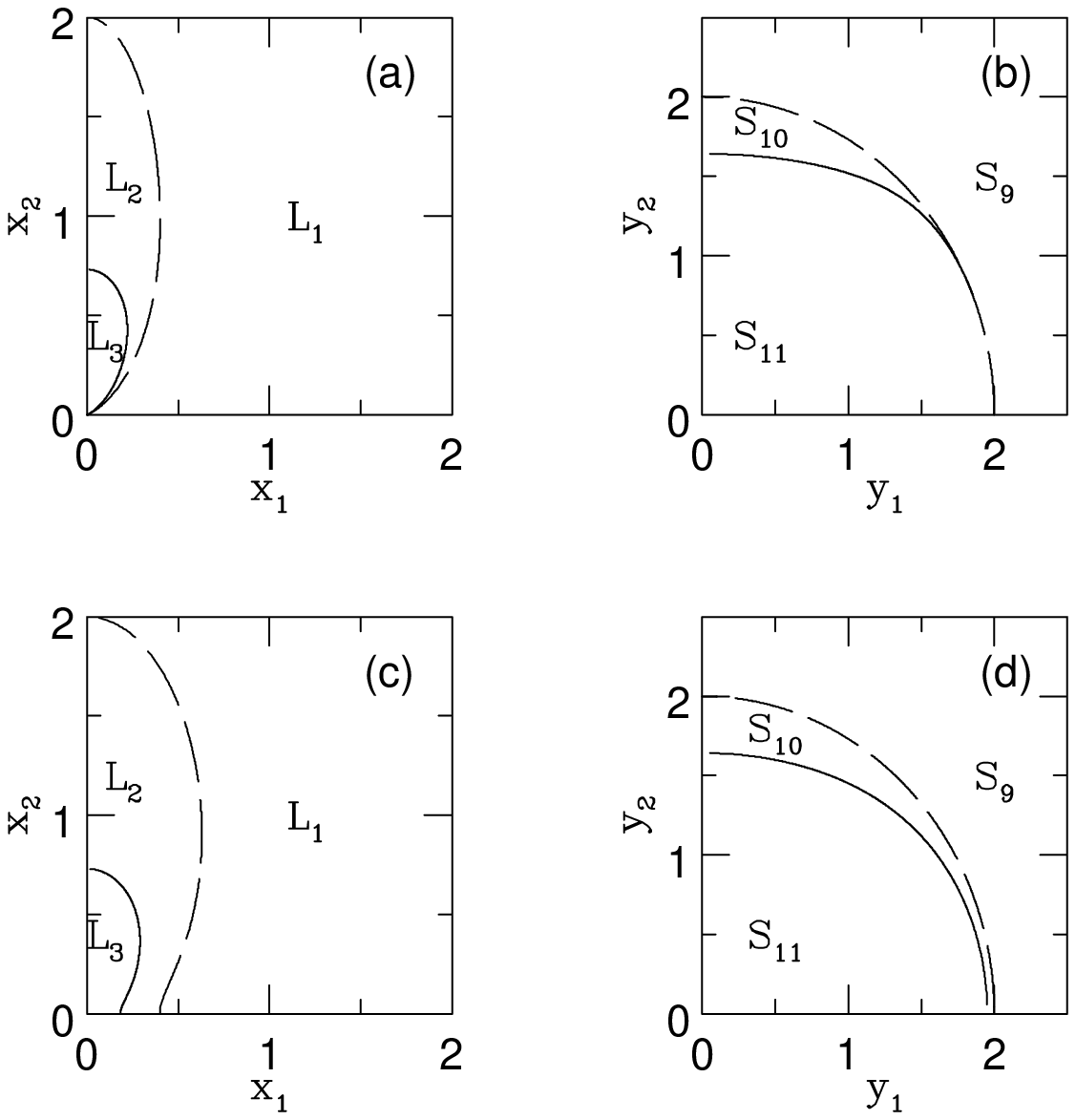,width=\hssize}

\smallskip\noindent
\caption{{\bf Figure 7.} The critical curve and caustic topologies
for the isothermal double power-law model perturbed with external
shear (continued). In panels (a) and (b), $0 > C_1 >-0.5 > C_2$ and
there is an F point on the $x_2$-axis and an S point at the origin; in
(c) and (d), $-0.5 > C_1> C_2$ and there is one F point on each axis.
The caustics and critical curves are drawn in full lines, the
pseudo-caustics and pseudo-critical curves are drawn in broken lines.
(Units used with $M=1$.)}

\endfigure

\beginfigure{8}

\psfig{figure=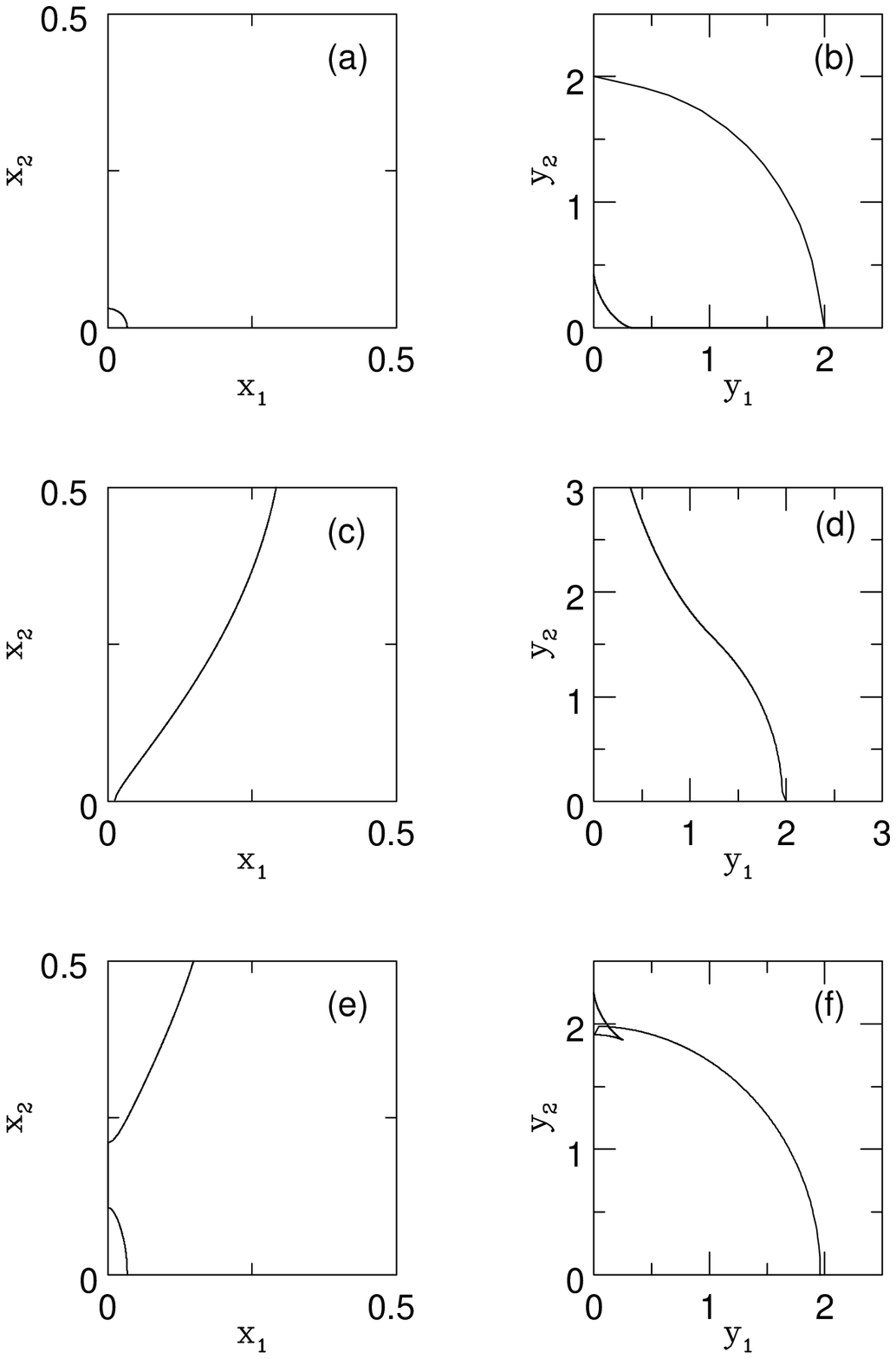,width=\hssize}

\smallskip\noindent
\caption{{\bf Figure 8.} As Fig.~6 but for an isothermal double 
power-law model with tiny core.  This causes the replacement of the
pseudo-caustics with true caustics and the destruction of the
pseudo-critical curves.  In each case, the central portion of the lens
plane is shown on the left, the source plane on the right. In panels
(a) and (b), $C_1 > C_2 > 0$ and there is one C and one F point on
each axis. Only the tiny inner radial critical curve is visible in the
plot. The outer tangential critical curve is virtually unchanged from
Fig.~6. In panels (c) and (d), $C_1 >0, 0> C_2 >-0.5$ and there is one
C point on the $x_2$-axis and one F point on the $x_1$-axis. In panels
(e) and (f), $C_1 >0, C_2 < -0.5$ and there is one F point on the
$x_1$ axis and a C point and two F points on the $x_2$-axis. (Units
used with $M=1$.)}

\endfigure

\eqnumber =1
\def\chaphead{\hbox{4.}}
\section{The Isothermal Double Power-Law Model}

One of the double power-law models has a remarkable property -- the
lens equation can be re-cast as a solvable, quadratic equation for the
image positions in terms of the source position. This is the model
with an isothermal cusp.  This property of invertibility is
exceedingly scarce. It is possessed by only two other known models --
the Schwarszchild lens (Einstein 1936; Refsdal 1964) and the infinite
isothermal sphere (Kovner 1987a).

\subsection{The Image Positions}

\noindent
When $\gamma =1$, the convergence diverges like $R^{-1}$ at small
radii and like $R^{-3}$ at large radii, viz;
\eqnam{\isothermal}
$$\kappa (R)= {M\over R(1+ R)^2},\qquad \psi = 2M \log ( 1 + R ).\eqno\new$$
Deprojecting the surface density by the well-known Abel inversion
formula (e.g., Binney \& Tremaine 1987), we find that the spherical
cluster has three-dimensional density
$$\rho (r)=\cases{ \displaystyle 
{\mtot \over 2\pi^2}\Bigl[ {2r^2+1\over (1-r^2)^2 r^2}
-{3 \arcosh (1/r)\over (1-r^2)^{5/2}}\Bigr],&$r\le 1$,\cr
\displaystyle 
 {\mtot \over 2\pi^2} \Bigl[ {2r^2+1\over (r^2-1)^2 r^2}
-{3 \arcos (1/r)\over (r^2-1)^{5/2}}\Bigr],&$r\ge1$.\cr}
\eqno\new$$
Here, $r$ is the three-dimensional radius in units of the length
scale.  A careful Taylor expansion shows the density is regular at $r
=1$ and has the value:
$$\rho(1) = {\mtot \over 5\pi^2}.\eqno\new$$
The three-dimensional density diverges like $r^{-2}$ at small
radii and like $r^{-4}$ at large radii. The model is therefore akin to
the famous halo model of Jaffe (1983).  We shall refer to it as {\it
the isothermal double power-law model}.  The deflection angle at a
displacement $x$ is
$$\Theta = \nabla \psi = {2M\sign (x) \over 1 + |x|}.\eqno\new$$
The circle $|y| = 2M$ is the pseudo-caustic.  Source positions with
$|y| < 2M$ give two images whereas those with $|y| > 2M$ give only
one.  The additional image is created or destroyed at the cusp.  The
pseudo-critical curves have degenerated into a point at the centre.

Taking $y>0$ without loss of generality, the images are located on
either side of the source at
$$x = \fr12 (y\mp 1) \pm \fr12 A_\pm (y),\eqno\new$$
with
$$A_{\pm}(y) = \sqrt{ (y \pm 1)^2 + 8M },\eqno\new$$
provided the source lies within the pseudo-caustic. The image at $x
>0$ is of type I, the image at $x <0$ is of type II.  If the source
lies outside the pseudo-caustic, then only the former image is
present.  When $|y| > 2M$, the total magnification is
$$\mu = {1\over 2} + {1\over 2y} {y(y+1) + 4M \over
                                   A_+(y)},\eqno\new$$
whereas when $|y| < 2M$, it is
$$\mu =  {1\over 2y} \Bigl[{y(y+1) + 4M \over
                           A_+(y)} + {y(y-1) + 4M \over
                           A_-(y)}\Bigr].\eqno\new$$
The time delay $\tau$ is also explicit. For any source position which
generates two images, it is
$$\eqalign{c\tau =& { \xic^2 \ds(1+\zd) \over \dd\dds}
                   \Bigl[ \fr14 (y+1)A_+(y) + \fr14 (y-1)A_-(y) \cr
                 & -y + 2M \log \Bigl|{ 1+ y + A_+(y)
                    \over  1 - y + A_-(y)}\Bigr|\Bigr],\cr}\eqno\new$$
where $\zd$ is the redshift of the lens. Fig.~5 shows the variation of
the time delay and the flux ratio of the images with separation of the
source $y$ from exact alignment.  For the images to be of comparable
brightness, we require $y \lta 1$. The time delay is then approximately
$$c\tau \approx {\sqrt{2M} \xic^2 \ds(1+\zd) \over \dd\dds},\eqno\new$$
and so scales roughly like $M^{1/2}$.

\subsection{Caustic Topology of the Perturbed Lens}

\noindent
The aim of this section is to classify the structure of the caustics
and critical lines of the perturbed isothermal model exactly. This job
has already been performed for the Plummer model (the instance $\gamma
=0$ in (\neatfamily)) in chapter 8 of Schneider, Ehlers \& Falco
(1992). The isothermal double power-law model has a pseudo-caustic not
present in the Plummer model. The slightest breakage of the circular
symmetry causes the pseudo-critical curves to be displaced from the
density singularity. These unusual features mean that it is well-worth
classifying the caustic topology for this model.

The lens action is considered as taking place in the presence of
an external larger-scale gravitational field with local surface mass
density $\fr12 \Gamma_1 + \fr12 \Gamma_2$ and local shear $\fr12
\Gamma_1 - \fr12\Gamma_2$.  By choosing the orientation of the
coordinates to diagonalise the tidal field, the lens equation becomes
$${\u y} = {\u x} \Bigl[ 1 - { 2M\over |{\u x}| (1 + |{\u x}|)
}\Bigr] - \left( \matrix{\Gamma_1&0\cr
           0&\Gamma_2\cr} \right) {\u x}.\eqno\new$$
Even in the presence of shear, the circle $|y| = 2M$ remains as
the pseudo-caustic. The pseudo-critical curves are now 
$$\cos^2 \theta = {4M(1+\Gamma_2) - r( 1 + \Gamma_2[\Gamma_2-2])
                  \over (\Gamma_1 - \Gamma_2) (r[\Gamma_1 + \Gamma_2
                  -2] - 4M)},\eqno\new$$
where ($r,\theta$) are polar coordinates in the lens plane.

Let us introduce the shorthand
$$C_1 = {M \over 1 - \Gamma_1}, \qquad C_2 = {M\over 1 - \Gamma_2}.
\eqno\new$$
Considering the first quadrant (i.e., $x_1 >0, x_2 >0$), critical 
points (C points) on the $x_1$-axis occur at
$$x_1 = \fr12 ( 1 + 8C_2)^{\fr12} - \fr12.\eqno\new$$
These exist provided $C_2 >0$. Folds (F points) occur at
$$x_1 = \sqrt{-2C_1} -1,\eqno\new$$
provided $C_1 < -\fr12$. Alternatively, the fold can occur at the
density singularity itself, when we refer to it as an S point. The
discussion for the $x_2$-axis is similar with trivial changes of
indices.

Figures 6 and 7 show the topologies of both the lens ($x_1,x_2$) and
source ($y_1,y_2$) planes for the distinct image configurations. In
Figs~6 (a) and (b), there is one C point on each axis. In Table 2, the
image regions corresponding to the different regions of the source
plane are listed. For example, a source in region S$_3$ produces four
images, one in L$_2$ in the first quadrant (L$_2\rone$), one in L$_2$
in the second quadrant (L$_2\rtwo$) and two in L$_3$ in the second and
third quadrants (L$_3\rtwo$, L$_3\rthree$). Moving the source from
S$_3$ to S$_2$ causes the two images in the second quadrant to fuse
and only two images are left, one in the first quadrant (L$_2\rone$)
and one in the third (L$_3\rthree$). If the source is now moved across
the pseudo-caustic from S$_2$ to S$_1$, the image in the third
quadrant vanishes into the central density singularity, while the
image in the first quadrant (L$_1\rone$) remains. As expected, when
the source crosses the pseudo-caustic, the number of images changes by
one.  Regions L$_1$, L$_2$ and L$_3$ correspond to images of types I,I
and II respectively. There is no change in parity on crossing the
pseudo-caustic.  In Figs~6 (c) and (d), there is a C point on the
$x_2$-axis and an S point at the origin. Whenever a critical curve
ends at an S point, the corresponding caustic is discontinuous.  As is
evident in the diagram, the caustic terminates abruptly in the source
plane as it touches the pseudo-caustic. Sources in S$_6$ generate
three images (see Table 2). One image is lost on crossing the
pseudo-caustic from S$_6$ to S$_5$, whereas one image is gained on
crossing the pseudo-caustic from S$_4$ to S$_5$. Regions L$_1$, L$_2$
and L$_3$ correspond to images of types II, II and III respectively.
In Figs~6 (e) and (f), there is both a C point and an F point on the
$x_2$-axis. Sources in S$_8$ have four images (see Table
2). Traversing the pseudo-caustic from S$_6$ to S$_8$ or from S$_4$ to
S$_7$ causes the gain of a single image. Regions L$_1$,L$_2$,L$_3$ and
L$_4$ correspond to images of types II,II,III and III respectively.
If $0>C_1>C_2>-0.5$ there are no critical points on the axes. Sources
with $|y| > 2M$ possess one image, source with $|y| < 2M$ possess no
images. The pseudo-critical curve lies at the origin.

The remaining two possible topologies are shown in Fig.~7. They are
included for completeness, but are not physical, as there are regions
of the source plane that correspond to no images in the lens plane. At
first sight, this seems to contravene the theorem of Schneider
(1984). If the time delay increases quadratically far from the lens,
it seems that there must be at least one minimum of the Fermat
surface. The reason why this theorem no longer holds here is that the
values of the shear $\Gamma_1$ and $\Gamma_2$ now exceed unity. The
Fermat potential decreases quadratically far from the lens and so
there must be a highest point on the Fermat surface -- but this
corresponds to the density spike and not an image. A somewhat similar
occurrence takes place in the singular point mass with shear or
Chang-Refsdal lens (see Schneider, Ehlers \& Falco, 1992, p. 261). In
Figs~7 (a) and (b), there is an F point on the $x_1$-axis and an S
point at the origin. Figs~7 (c) and (d) show a similar configuration,
but with an F point on each axis. In both cases, sources in S$_9$ give
rise to one image of type III in region L$_1$.  Crossing the
pseudo-caustic from S$_9$ to S$_{10}$ generates a second image of type
II in L$_3$, while the pre-exisiting image moves from L$_1$ to
L$_2$. However, as the source crosses the caustic to S$_{11}$, both
images merge and disappear into the central density singularity. Note
that in Fig.~7 (b), the crossing of the pseudo-caustic from S$_9$ to
S$_{11}$ is accompanied by the destruction of a single image whereas
the crossing from S$_9$ to S$_{10}$ is accompanied by the creation of
a single image.  Sources in S$_{11}$ have no images at all.

As an aid to understanding, it is helpful to introduce a tiny core
into the isothermal double power-law model so that it becomes
$$\kappa = {M\over (R+\epsilon)(1+ R)^2},\eqno\new$$
with corresponding deflection angle
$$\Theta = {2M\sign (x) \over (1 - \epsilon) (1 + |x|)}
           -{2M\epsilon \over |x| (1-\epsilon)^2}\log\Bigl[
           {|x| + \epsilon \over \epsilon (1 + |x|)}\Bigr].\eqno\new$$
The changes in the topology of the caustics and the critical curves
are shown in detail in Fig~8. This figure is directly comparable with
the cases examined in Fig~6.  The introduction of the infinitesimal
core causes the conversion of the pseudo-caustics into true caustics,
as is evident on comparing Figs~8 (b) and (f) with Figs~6 (b) and
(f). Now, the number of images changes by two on crossing these
curves. The pseudo-critical curves are completely destroyed by the
introduction of the core. Replacing them are tiny critical curves
generated near the origin. For example, Figs~8 (a) and (e) show the
emergence of a minute radial critical curve from the origin because of
the infinitesimal core. Only the inner portion of the lens plane is
shown in the figures -- the outer critical curve is virtually
unchanged from the singular model in Figs~6 and 7. So, not all the
critical points on axis are visible in the details of the central
regions. In Fig.~8 (d), only part of the pseudo-caustic is converted
to a caustic, the remaining part being destroyed. In the lens plane,
the S point of Fig.~6 (c) becomes an F point in Fig.~8 (c).

\eqnumber =1
\def\chaphead{\hbox{5.}}
\section{The Power-Law Lenses} 

In this section, we study density cusps of infinite extent with
elliptic equipotentials.  Lenses with equipotentials stratified on
similar concentric ellipses have been studied before (e.g., Kovner
1987b; Blandford \& Kochanek 1987; Kochanek \& Blandford 1987;
Kassiola \& Kovner 1993; Witt 1996; Witt \& Mao 1997), although the
emphasis has generally been on models with softened cores. By
contrast, our interest is focused on the lenses with central density
singularities.
 
\subsection{The Potential-Convergence Pair}

The deflection potential is
\eqnam\powerpot
$$\psi = A[ x_1^2 + x_2^2 \qm2 ]^{1 - \gamma/2},\eqno\new$$ 
where $q$ is the axis ratio of the equipotentials, $\gamma$ is the
cusp index and the constant $A$ fixes the overall scaling. All the
models have infinite total projected mass.

\beginfigure{9}

\centerline{\psfig{figure=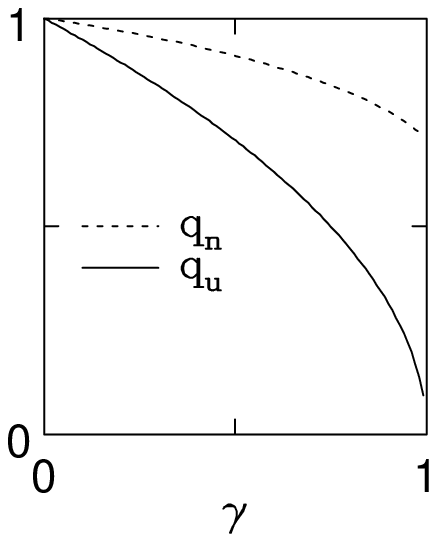,width=0.4\hssize}}

\smallskip\noindent
\caption{{\bf Figure 9.} The critical values of q for naked cusps (q$_n$) 
and the umbilic catastrophe (q$_u$) plotted as functions of $\gamma$}

\endfigure

As Contopoulos (1954) noted, three-dimensional distributions that are
stratified on similar concentric ellipsoids project to two-dimensional
distributions stratified on similar concentric ellipses. The projected
potential (\powerpot) therefore corresponds to the three-dimensional
galaxy models written down by Evans (1993, 1994) and called the
power-law galaxies. Schneider, Ehlers \& Falco (1992) caution that
models with elliptical equipotentials may not be very realistic, so it
is worth pointing out that the power-law galaxies have been
successfully used to represent the nearby elliptical M32 (van der
Marel et al. 1994) as well as the inner 500 parsecs of the Galactic
bulge (Evans \& de Zeeuw 1994).  The convergence is cusped like
$$\kappa = {A(2-\gamma)\over 2q^2}{ax_1^2 + bx_2^2 \over 
(x_1^2 + x_2^2 \qm2)^{1 + \gamma/2}},\eqno\new$$
with 
$$a= 1 - q^2(1-\gamma),\quad b = 1 - \qm2 (1-\gamma).\eqno\new$$
The axis ratio of the equipotentials $q$ describes the flattening of
the model and can be restricted to lie in the range $0 \le q \le 1$
without loss of generality.  The projected mass density of the
singular models has ellipticity $\epsilon$
$$\epsilon = 1 - q \Bigl[ {q^2-\gamma +1 \over 1 + q^2(1-\gamma)}
                       \Bigr]^{1/\gamma}.\eqno\new$$
The convergence is positive if $\gamma < 1 + q^2$ (e.g., Blandford \&
Kochanek 1987). If the model is viewed as a projected power-law
galaxy, the constraints that the three-dimensional density and that
the two-integral distribution function are positive definite are more
severe and given in Evans (1994). The parameter $\gamma$ describes the
rotation curve of the model. If $\gamma = 1$, the model has an
asymptotically flat rotation curve (e.g., Binney \& Tremaine 1987;
Evans 1993). If $\gamma >1$, the rotation curve is declining, whereas
if $\gamma <1$, the rotation curve is rising. The range of $\gamma$ is
normally restricted to $0< \gamma <2$. The lower limit comes from
requiring the convergence to vanish at spatial infinity, the upper
limit is the Keplerian case of a point mass.

The tangential critical curve has intercepts
$$\xonet^2 = \Bigl[ {Aq^{-2} (2 - \gamma)} \Bigr]^{2/\gamma},
\eqno\new$$
$$\xtwot^2 = q^2 \Bigl[ A(2-\gamma) \Bigr]^{2/\gamma}.\eqno\new$$
The condition for the existence of a tangential critical curve is
$$A(2-\gamma) > 0.\eqno\new$$
It is always satisfied. If the cusp is strong ($\gamma > 1$), there is
a transition from two to four images on crossing the tangential
caustic. If the cusp is weak ($\gamma < 1$), there is a transition
from one to three images. The radial critical curve has intercepts
$$\xoner^2 = \Bigl[ A(2-\gamma)(1-\gamma)\Bigr]^{2/\gamma}.\eqno\new$$
$$\xtwor^2 = q^2 \Bigl[A q^{-2}(2-\gamma)(1-\gamma) \Bigr]^{2/\gamma}.
\eqno\new$$
The radial critical curve exists only if $\gamma <1$. Quintuple imaging
occurs if
$$A(2-\gamma)(1-\gamma) > 0.\eqno\new$$
That is, quintuple imaging can occur for all power-law galaxies with weak
cusps. The critical curves become smaller both in the
limit of increasing homogeneity ($\gamma \rightarrow 0$) and in the
Keplerian point mass limit ($\gamma \rightarrow 2$).

In the source plane, the semi-axes of the tangential and radial
caustics are
$${\ytwot\over \yonet} = {\yoner\over \ytwor}= q^{2/\gamma -1}.\eqno\new$$
For $\gamma < 1$, there are two important topology changes that can
occur. For models only weakly perturbed from circular symmetry, the
radial caustic lies entirely outside the tangential caustic.  Models
flatter than the critical flattening $\qn$ given by
\eqnam{\critnaked}
$$\qn^2 = 1 - \qn^{2/\gamma} \gamma (1 - \gamma)^{1/\gamma -1},\eqno\new$$
possess \lq\lq naked cusps''. That is, two of the cusps of the
tangential caustic lie outside the radial caustic. The umbilic
catastrophe occurs at a critical flattening $\qu$ given by
\eqnam{\crituc}
$$\qu^2 = 1 - \gamma.\eqno\new$$
Fig.~9  shows the behaviour of the critical axis ratio $\qn$ and
$\qu$ for the onset of naked cusps and the umbilic catastrophe
respectively as a function of cusp index $\gamma$.

\beginfigure{10}
\psfig{figure=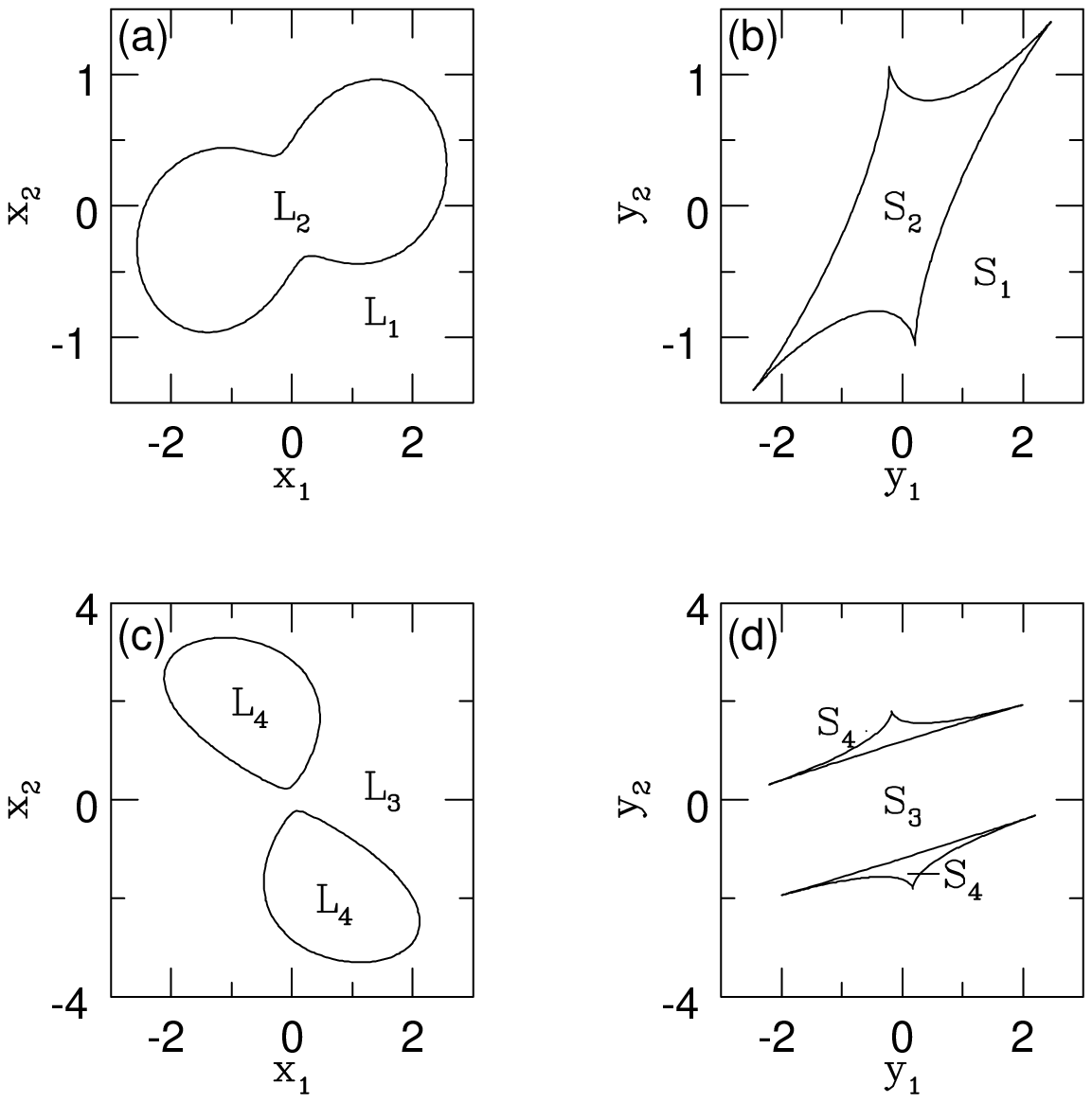,width=\hssize}
\smallskip\noindent
\caption{{\bf Figure 10.} The critical curve and caustic topologies
for the power-law lenses in the presence of external shear. In each
case, the lens plane is shown on the left, the source plane on the
right.  This figure is for the case of a strong cusp with $\gamma =
1.25$ and $q =0.6$. [Figs~10 (a) and (b) have $\Gamma_1 = 0.1,
\Gamma_2 = 0.5$, Figs~10 (c) and (d) have $\Gamma_1 = 0.8, \Gamma_2 =
0.7$]}

\endfigure

\beginfigure{11}
\psfig{figure=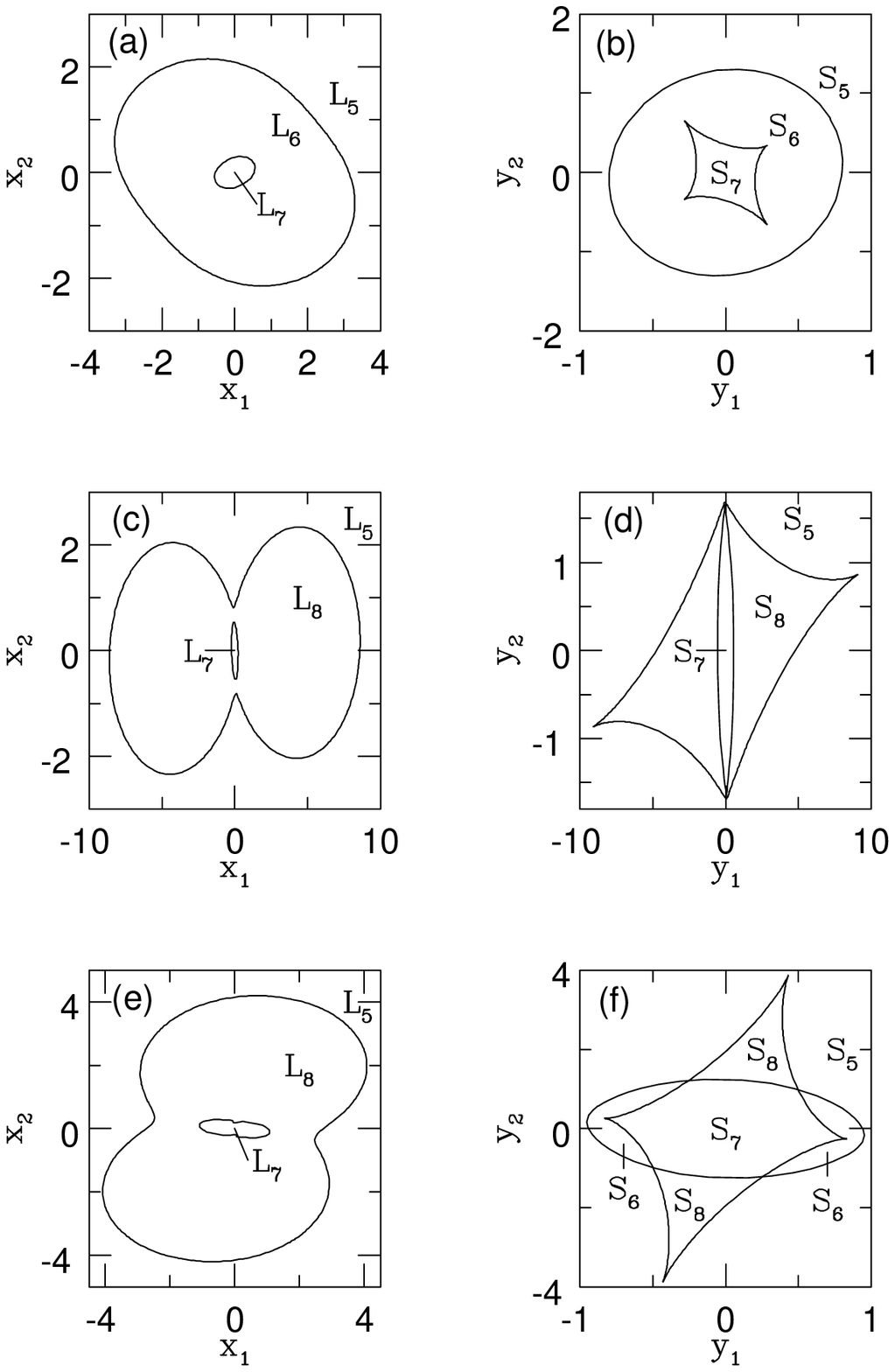,width=\hssize}
\smallskip\noindent
\caption{{\bf Figure 11.} As in Figure 10, but for the case of a
weakly cusped power-law lens with $\gamma = 0.75$ and $q =0.6$. [Figs~11
(a) and (b) have $\Gamma_1 = -0.5, \Gamma_2 = -0.1$, Figs~11 (c) and
(d) have $\Gamma_1 = 0.3, \Gamma_2 = 0.1$ and Figs~11 (e) and (f) have
$\Gamma_1 = -0.7, \Gamma_2 = 0.1$]}
\endfigure

For the moment, let us consider the model without external shear. In
general, the lens equation must be solved numerically. However, when
the source is on-axis and aligned with the lens, the image positions
and magnifications can be found exactly.  This is sufficiently
uncommon for non-axisymmetric lenses that it is worth giving the
solution explicitly. There are always two images on the minor axis at
$(0,\pm x_2)$, where
$$x_2 =q \Bigl[ {A(2-\gamma)\over q^2} \Bigr]^{1/\gamma}.\eqno\new$$  
These have even parity and the magnification $\mu$ is
$$\mu = {1\over \gamma (1 - q^2)}.\eqno\new$$
There are always two images on the major axis at ($\pm x_1,0$), where
$$x_1 = \Bigl[{A(2-\gamma)}\Bigr]^{1/\gamma}.\eqno\new$$  
These have odd parity and the magnification $\mu$ is
$$\mu = {q^2\over \gamma (1 - q^2)}.\eqno\new$$
Only these four images occur when the cusp is strong. For weak cusps,
there is an additional (infinitely de-magnified) image of even parity
at the origin. The total magnification $A_{\rm tot}$ is
$$A_{\rm tot} = {2(1+q^2)\over \gamma (1-q^2)}.\eqno\new$$
This equation is exact when the source is on-axis and is approximate
when the source lies within the tangential caustic. This equation has
previously been given by Kassiola \& Kovner (1993) for the special
case $\gamma =1$.

The $\gamma =1$ model has the characteristic properties of an
isothermal cusp.  Unlike the perturbed circular model of Section 4.2,
it has a truly non-axisymmetric pseudo-caustic. Its equation can be
calculated by the methods of Section 2.2.  If the phase of $z =x_1 + i
x_2$ is $\theta$ and the phase of $\zs = y_1 + i y_2$ is $\phi$, then
the equation of the pseudo-caustic is
$$ \eqalign{|\zs|^2 =& y_1^2 + y_2^2 =
 \Bigl({\partial \psi \over \partial x_1}\Bigr)^2 
            + \Bigl({\partial \psi \over \partial x_2}\Bigr)^2 \cr
            =& {A^2 (\cos^2\theta + q^{-4} \sin^2\theta)
\over \cos^2\theta + \qm2 \sin^2 \theta}.\cr}\eqno\new$$
As the source crosses the pseudo-caustic, an extra image of type II is
generated at the density singularity. On the pseudo-caustic, we know
from the lens equation that the relation between the two phases is
$$\tan \theta = \qm2 \tan \phi, \eqno\new$$
and so the pseudo-caustic is an ellipse of form
\eqnam{\pseudoc}
$$y_1^2 + q^2y_2^2 = A^2.\eqno\new$$
The pseudo-caustic is concentric with the elliptic equipotentials, but
it is elongated in the opposite direction. For this model, the lens
equation can be reduced to a quartic equation. The quartic degenerates
to simpler quadratics if the source lies either on the major axis or
on the minor axis. Suppose the source lies on the major axis at
$(y_1,0)$, with $y_1 >0$ without loss of generality. Then there is one
image if $y_1 > A$ at
\eqnam{\imagechange}
$$x_1 = y_1 +A, \qquad x_2 =0.\eqno\new$$
Initially, this image has even parity. Its magnification is
$$\mu = {A + y_1 \over |y_1 - A(\qm2 -1)|}.\eqno\new$$
On crossing the pseudo-caustic at $y_1 = A$, there is a second image
generated at 
$$x_1 = y_1-A, \qquad x_2=0.\eqno\new$$
It has odd parity and finite magnification
$$\mu = {A - y_1 \over A(\qm2 -1) + y_1}.\eqno\new$$
On crossing the tangential caustic at $y_1 = A(\qm2-1)$, the image
(\imagechange) changes from even to odd parity.  Simultaneously, two
new images of even parity are generated at
$$x_1 = {y_1\over (1-q^2)}, \qquad x_2 = \pm {1\over q}\Bigl[A^2 - 
                  {y_1^2\over (\qm2-1)^2 }\Bigr]^{1/2}.\eqno\new$$
The magnification of each image is
$$\mu = {A^2\qm2 (\qm2 -1) \over A^2(\qm2 -1)^2 - y_1^2}.\eqno\new$$
Note that a change in parity of one of the pre-existing images must
occur on crossing a pseudo-caustic for our general theorems of Section
2.2 to hold good. The images and magnifications when the source lies on
the minor axis can be found similarly.

\subsection{Caustic Topology of the Perturbed Lens}

\begintable{3}%
\caption{{\bf Table 3.} Image Positions as Function of Source
Positions as Labelled on Figs 10 and 11. }
\halign{#\hfil&\quad#\hfil\cr
\noalign{\hrule}
\noalign{\vskip0.3truecm}
S$_1$         &L$_1$, L$_2$  \cr
\noalign{\vskip0.2truecm}
S$_2$         &2L$_1$, 2L$_2$  \cr
\noalign{\vskip0.2truecm}
S$_3$         &2L$_3$  \cr
\noalign{\vskip0.2truecm}
S$_4$         &3L$_3$, L$_4$ \cr
\noalign{\vskip0.2truecm}
S$_5$         &L$_5$ \cr
\noalign{\vskip0.2truecm}
S$_6$         &L$_5$, L$_6$, L$_7\,\,$ \cr
\noalign{\vskip0.2truecm}
S$_7$         &2L$_5$, 2L$_6$, L$_7\,\,$ \cr
\noalign{\vskip0.2truecm}
S$_8$         &2L$_5$, L$_8$ \cr
\noalign{\vskip0.1truecm}
\noalign{\hrule}
}
\endtable

\noindent
Figures 10 and 11 show the possible topologies of the caustics for the
power-law lenses in the presence of external shear. In this case, the
lens equations become
$$ y_1 = x_1\Bigl[1 + \Gamma_1 - {A(2-\gamma)\over (x_1^2 +
x_2^2q^{-2})^{\gamma/2}}\Bigr] + \Gamma_2x_2,\eqno\new$$
and
$$ y_2 = x_2\Bigl[1 - \Gamma_1 - {Aq^{-2}(2-\gamma)\over (x_1^2 +
x_2^2q^{-2})^{\gamma/2}}\Bigr] + \Gamma_2x_1,\eqno\new$$
where the shear matrix is taken as traceless and of form:
$$\Gamma = \left( \matrix{\Gamma_1&\Gamma_2\cr
           \Gamma_2&-\Gamma_1\cr} \right).\eqno\new$$
These equations have been recently considered by Witt \& Mao (1997),
who prove a number of interesting theorems regarding the image
positions.  Since the shear axes do not, in general, coincide with the
major and minor axes of the lens potential, the symmetry axes of the
caustics are also different from those of the lens. Consequently,
Table 3 -- which gives the image regions corresponding to different
regions in the source plane -- does not contain information about
which quadrants will contain the images. The quadrant in which a given
image is found varies according to the exact position of the source.

Fig.~10 shows the two distinct topologies which occur for a strong
cusp ($\gamma = 1.25$). When the external shear is small compared to
the lens shear, there is only one critical curve in the lens plane
and a corresponding caustic in the source plane, as shown in Figs~10
(a) and (b).  A source in region S$_2$ produces four images, two in region
L$_1$ of the lens plane and two in region L$_2$. Moving the source
from S$_2$ to S$_1$ causes two of these images to fuse together
leaving only two images, one in region L$_1$ and the other in region
L$_2$.  If $\Gamma_1$ and $\Gamma_2$ satisfy
\eqnam{\gamcircle}
$$ \Gamma_1^2 + \Gamma_2^2 = 1,\eqno\new$$
then the caustic becomes extremely elongated along the axis
$$y_2 = {\Gamma_2 \over 1 + \Gamma_1}.\eqno\new$$
The result of further increasing the shear so that $\Gamma_1^2 +
\Gamma_2^2 > 1$ is shown in Figs~10 (c) and (d) -- there are now two
critical curves and two caustics. A source in region S$_4$ produces
four images, three in region L$_3$ and one in region L$_4$. If the
source if moved from S$_4$ to S$_3$, two of the images fuse and only
two images are left, both lying in region L$_3$.  It is also possible
to have a situation in which there is no caustic.

%
%
%
%

Fig.~11 shows the three possible critical curve and caustic topologies
for the case of a weak cusp ($\gamma = 0.75$).  Figs~11 (a) and (b)
show the radial caustic lying entirely outside the tangential
caustic. Alternatively, as in Figs~11 (c) and (d), the tangential
critical curve can completely contain the radial caustic.  For
intermediate values of the shear, the tangential curve can lie partly
outside the radial curve and partly inside, as shown in Figs~11 (e)
and (f).  If we continue to increase the magnitude of the external
shear, the outer curve in Fig.~11 (d) is observed to be stretched
along its major axis while being simultaneously compressed along its
minor axis. When $\Gamma_1$ and $\Gamma_2$ satisfy (\gamcircle), this
curve becomes a straight line. Outside the unit circle in the
($\Gamma_1,\Gamma_2$) plane, there exists only one caustic, with two
cusps.


In Figs.~11 (a) and (b), a source placed in region S$_7$ produces five
images as expected for a weak cusp. There are two in region L$_5$, two
in L$_6$ and one in L$_7$. If the source is moved across the caustic
to region S$_6$, one pair of images fuses together to leave three
images, one in each of the regions L$_5$, L$_6$ and L$_7$. Finally, if
the source is moved to region S$_5$ another pair of images fuses to
leave a single image in region L$_5$. The locations of the images
corresponding to the source regions in Figs.~11 (d) and (f) can be
found from Table 3. As described above, if $\Gamma_1$ and $\Gamma_2$
lie outside the unit circle in the ($\Gamma_1$,$\Gamma_2$) plane, the
tangential critical curve no longer exists. Sources placed inside the
one remaining caustic give three images (two lie outside the critical
curve and one lies inside) while those placed outside give only one
image lying outside the critical curve.

\eqnumber =1
\def\chaphead{\hbox{6.}}
\section{Conclusions} 

This paper has studied the properties of lenses with central density
cusps. These are models in which the convergence $\kappa$ diverges as a
function of projected radius $R$ like 
$$\kappa \propto R^{-\gamma},\eqno\new$$
near the centre. Numerical simulations (e.g., Navarro, Frenk \& White
1996), dynamical arguments (e.g., Evans \& Collett 1997) and
observations (e.g., Lauer et al. 1995; Faber et al. 1997) all suggest
that galaxies and clusters commonly have such singular density
profiles. By generalising the index theorem to allow for central
density singularities, we suggest the following classification of
cusps on the basis of their lensing properties:

\medskip
\noindent
(1) Strong density cusps ($1 < \gamma <2$) give rise to an even number
of images.  There are as many images of odd parity as there are images
of even parity ($\ni - \nii + \niii =0$). Models with strong cusps do
not possess radial caustics.

\medskip
\noindent
(2) Weak density cusps ($0 < \gamma <1$) give rise to an odd number of
images.  The images of even parity outnumber the images of odd parity
by one ($\ni - \nii + \niii =1$). Models with weak cusps possess
radial caustics. Even though such models are singular, they can
produce radial arcs (cf. Mellier, Fort \& Kneib 1993; Bartelmann
1996).

\medskip
\noindent
(3) Isothermal density cusps ($\gamma =1$) always possess {\it
pseudo-caustics}.  If the source is within the pseudo-caustic, there
is an even number of images ($\ni - \nii + \nii =0$).  If the source
is outside the pseudo-caustic, there is an odd number ($\ni - \nii +
\nii =1$). A method for computing the location of the pseudo-caustic
is given. Note that the pseudo-caustic differs from a true caustic in
two ways. First, the magnification of a point source is finite, not
infinite, at the pseudo-caustic.  Second, the number of images created
or destroyed on crossing the pseudo-caustic is one, not two. A
circular lens, even if perturbed by external shear, has a
pseudo-caustic that is a circle. This is not the case if the lens is
truly non-axisymmetric.

\medskip
As specific examples, we have presented two families of models. The
double power-law family has a central cusp, a transition region and an
outer envelope. The power-law family are a set of elliptical cusps of
infinite extent. They are projected power-law galaxies. In both cases,
detailed properties of the individual lenses are given, together with
a classification of the caustic structures in the presence of external
shear.  As an application, we consider a problem pointed out by
Bartelmann (1996). He suggested that lensing may be problematic for
the cusped Navarro, Frenk \& White (1996) density profile in cases
where both radial and tangential arcs are present. On matching the
positions of the arcs, this model predicts large values for the
radial magnification at the tangential arc, which in turn implies that
the corresponding sources must be radially very thin. We have shown
that this problem is not so serious for lenses that have more singular
density cusps -- and, indeed, higher resolution simulations of halo
formation suggest that the cusp slope may have been originally
underestimated (Fukushige \& Makino 1997; Moore et al 1997).

Our investigation has uncovered a new model -- {\it the isothermal
double power-law model} -- for which the lens equation is exactly
solvable for any source position. This scarce property is only
possessed by two other known circularly symmetric lenses, the
Schwarzschild lens and the isothermal sphere.  The isothermal double
power-law model has a three-dimensional density which behaves like
$r^{-2}$ in the centre and like $r^{-4}$ in the outer parts, so it is
a realistic model for a galaxy or cluster of finite extent with a flat
rotation curve. As the image positions are explicitly available, the
lensing cross-sections are straightforward to calculate.

\section*{Acknowledgments}
NWE thanks the Royal Society for financial support. MW acknowledges
financial support from a Scatcherd Scholarship. The paper was partly
written at the Aspen Center for Physics. We particularly wish to thank
Chris Kochanek, Donald Lynden-Bell and Prasenjit Saha for a number of
helpful conversations.

%
%
%

\bye